\documentclass[12pt,english,preprint,iop]{revtex4-1}

\usepackage{times}
\usepackage{array}
\usepackage{longtable}
\usepackage{float}
\usepackage{amsmath}
\usepackage{bm}
\usepackage{graphicx}
\usepackage{amssymb}
\usepackage{xcolor}
\usepackage{hyperref}
\hypersetup{bookmarks=true,colorlinks=true, linkcolor=blue,linktoc=all}
\usepackage[FIGTOPCAP]{subfigure}
\usepackage{ulem}

\makeatletter
\usepackage{fancyhdr}
\usepackage{babel}
\makeatother

\begin{document}

\title{A Runaway Electron Avalanche Surrogate for Partially Ionized Plasmas}

\author{Jonathan S. Arnaud}
\email{j.arnaud@ufl.edu}
\affiliation{Nuclear Engineering Program, Department of Materials Science and Engineering, University of Florida, Gainesville, FL 32611, United States of America}
\author{Xian-Zhu Tang}
\affiliation{Theoretical Division, Los Alamos National Laboratory, Los Alamos, NM 87545, United States of America}
\author{Christopher J. McDevitt}
\email{cmcdevitt@ufl.edu}
\affiliation{Nuclear Engineering Program, Department of Materials Science and Engineering, University of Florida, Gainesville, FL 32611, United States of America}

\date{\today}

\begin{abstract}
A physics-constrained deep learning surrogate that predicts the
exponential ``avalanche'' growth rate of runaway electrons (REs) for a plasma
containing partially ionized impurities is developed. Specifically, a
physics-informed neural network (PINN) that learns the adjoint of the
relativistic Fokker-Planck equation in steady-state is derived,
enabling a rapid surrogate of the RE avalanche for a broad range of
plasma parameters, motivating a path towards an ML-accelerated
integrated description of a tokamak disruption. A steady-state power
balance equation together with atomic physics data is embedded
directly into the PINN, thus limiting the PINN to train across
physically consistent temperatures and charge state
distributions. This restricted training domain enables accurate
predictions of the PINN while drastically reducing the computational
cost of training the model. In addition, a novel closure for the
relativistic electron population used when evaluating the secondary
source of REs is developed that enables improved accuracy compared to
a Rosenbluth-Putvinski source. The avalanche surrogate is verified
against Monte Carlo simulations, where it is shown to accurately
predict the RE avalanche growth rate across a broad range of plasma
parameters encompassing distinct tokamak disruption scenarios.
\end{abstract}

\maketitle
\section{Introduction}
A crucial issue in the continued development of the tokamak approach
to fusion energy is the impact of disruptions on plasma facing
components. A tokamak disruption consists of the loss of magnetic
confinement leading to the rapid collapse of the plasma
temperature. Following the loss of thermal energy, the plasma current
decays on a longer timescale, leading to the release of the plasma's
magnetic energy. The strong inductive electric field generated prior to
this current quench phase of a disruption often leads to the formation
of a relativistic electron population~\cite{hender2007mhd,
  breizman2019physics}. These so-called runaway electrons
(RE)~\cite{wilson1925acceleration} will often become sufficiently
numerous such that they are able to carry nearly all of the plasma
current~\cite{jayakumar1993collisional,rosenbluth1997theory}. The generated RE population will typically form a collimated beam of electrons and often times terminate on a highly localized region of the tokamak wall, thus creating a dangerous scenario for plasma-facing components on existing and future tokamaks~\cite{matthews2016melt}.

While existing experiments are actively pursuing mitigation scenarios and improving the understanding of tokamak disruptions, they cannot access the plasma parameters relevant to next generation tokamaks, motivating the need for high physics fidelity models. One ongoing challenge facing high physics fidelity models is the development of an integrated description of a tokamak disruption. In particular, an integrated description of a tokamak disruption must account for three distinct physical processes, which consist of a collisional radiative model describing radiative losses and charge state distributions, the formation and evolution of REs, and the magnetohydrodynamic (MHD) activity during the disruption. Integrating these processes in a self-consistent description of a disruption is challenging due to the computational demands of each model.

This work focuses on RE formation during a tokamak disruption. During the rapid collapse of the plasma electron temperature $T_e$, the resistivity ($\eta \propto T_e^{-3/2}$) increases significantly, which drives a sharp increase of the electric field ($E = \eta j$), where $j$ is the current density, providing a robust drive for RE generation~\cite{dreicer1959electron,connor1975relativistic,chiu1998fokker,harvey2000runaway,helander2004electron,smith2005runaway}. The initial population of REs, often referred to as `seed' runaway electrons, will then undergo large-angle Coulomb collisions with thermal electrons, leading to the formation of secondary runaway electrons, allowing for the RE population to grow exponentially~\cite{sokolov1979multiplication,jayakumar1993collisional,rosenbluth1997theory}. This process of exponentially generating REs (known as the RE avalanche) is expected to be the driving mechanism that enables REs to potentially carry the majority of the pre-disruption plasma current in reactor scale devices such as ITER. The RE population can obtain energies up to tens of mega-electron volts, thus having the ability to induce significant surface and subsurface damage to tokamak components. Due to the exponential sensitivity of the avalanche mechanism, a high-fidelity description is required to accurately predict the final amount of RE current at the end of a disruption.

One promising path to rapidly accelerate the description of REs is the use of deep learning methods to develop rapid surrogates of distinct RE mechanisms~\cite{hesslow2019evaluation,mcdevitt2023physics,yang2024pseudoreversible,arnaud2024physics,mcdevittpart12025,mcdevittpart22025}. Recently, a deep learning-based surrogate describing the avalanche generation of REs~\cite{arnaud2024physics} was developed, providing a proof-of-principle demonstration of this approach. This recently developed surrogate, however, did not include the impact of partial screening, which is known to strongly impact RE generation rates and thresholds~\cite{mosher1975interactions, martin2017formation, hesslow2017effect, hesslow2019influence, mcdevitt2019avalanche}. During a disruption, impurities are often injected or released from the wall, where the low temperatures of the post-thermal quench plasma result in a partially ionized plasma~\cite{lehnen2015disruptions}. Thus, a primary aim of this paper is to demonstrate the extension of the recently developed surrogate by including the effects of partial screening on RE avalanching.

While a surrogate of the RE avalanche for an arbitrary combination of plasma parameters may be useful for particular applications, certain plasma parameters are strongly correlated during a tokamak disruption. By accounting for such correlations, this can reduce the range of solutions the deep learning model is trained across, hence improving the robustness of the model while reducing the computational cost of training the model. As previously mentioned, the rapid collapse of the plasma temperature will lead to a corresponding electric field that will largely dictate the efficiency of the RE avalanche; however, the final plasma temperature after the initial cooling phase of the disruption will depend sensitively on the amount of impurities that are present in the plasma. Thus, by integrating a power balance equation, Ohm's law, and an impurity charge state computed from data generated by a collisional radiative model into the deep learning model, this allows us to infer the RE avalanche growth rate for a given plasma composition and current density. Such a surrogate, while containing increased physics complexity, is found to not pose a problem in development, thus motivating a path to developing high fidelity integrated descriptions of tokamak disruptions.

Additionally, the RE avalanche surrogate developed in Ref.~\cite{arnaud2024physics} utilized a Rosenbluth-Putvinski secondary source term~\cite{rosenbluth1997theory} to predict the avalanche growth rate, which assumes all REs to have infinite energy and to be aligned with the magnetic field. It was found that this approximation was susceptible to inaccuracies near the RE avalanche threshold, where modestly strong electric fields were considered (up to ten times the Connor-Hastie electric field~\cite{connor1975relativistic}). In this work we will increase the fidelity of the RE avalanche surrogate by utilizing a more accurate secondary source term~\cite{chiu1998fokker,harvey2000runaway}, that accounts for the energy distribution of REs. Since the energy distribution of REs is not predicted by the formulation employed here, we will take the REs to obey an ``avalanche'' distribution~\cite{rosenbluth1997theory}, which corresponds to an exponentially decaying distribution of REs with an `avalanche temperature' set by the avalanche growth rate. An iterative scheme is used to self-consistently evaluate the avalanche temperature, where the resulting prediction of the avalanche growth rate is shown to be substantially improved compared to those of Ref.~\cite{arnaud2024physics} that utilized the Rosenbluth-Putvinski secondary source term, even with the increased physics complexity and broader range of parameters considered in this paper. Upon acceptance of this manuscript, the PINNs developed in this paper will be uploaded to the public repository: \url{https://github.com/cmcdevitt2/RunAwayPINNs}, where additional ML surrogates for REs are available~\cite{mcdevitt2023physics,arnaud2024physics,mcdevittpart12025,mcdevittpart22025}.

The rest of this paper is organized as follows: Section~\ref{sec:PCDL} briefly describes the physics-constrained deep learning approach used to describe RE formation. In Sec.~\ref{sec:RPR} we demonstrate the performance of the physics-constrained deep learning RE description. Section~\ref{sec:SSRES} describes the formulation of the RE avalanche surrogate, with Sec.~\ref{sec:PIREAS} demonstrating the use of the surrogate to infer relevant quantities of interest. Conclusions and a brief discussion are given in Sec. \ref{sec:D}.

\section{\label{sec:PCDL}Physics-Constrained Deep Learning of the Adjoint to the Steady-State Fokker Planck Equation}
\subsection{Physics-informed neural networks}
We briefly describe the physics-constrained deep learning approach used in this paper, which involves the embedding of physical information during the training process of a machine learning (ML) model, enabling the development of rapid surrogate models [see Refs.~\cite{sun2020surrogate,haghighat2021physics} for examples]. The specific approach used here is a physics-informed neural network (PINN)~\cite{raissi2019physics}, which in the data-free limit embeds a partial differential equation (PDE), boundary conditions, and initial conditions directly into the loss function $\mathcal{L}$, i.e.
\begin{align}
\mathcal{L} = \frac{1}{N_{PDE}} &\sum^{N_{PDE}}_i  \mathcal{R}^2 \left( \mathbf{p}_i, t_i; \bm{\lambda}_i \right) + \frac{1}{N_{bdy}} \sum^{N_{bdy}}_i \left[ y_i - y \left( \mathbf{p}_i, t_i; \bm{\lambda}_i \right)\right]^2 \nonumber \\
+ \frac{1}{N_{init}} &\sum^{N_{init}}_i \left[ y_i - y \left( \mathbf{p}_i, t=0; \bm{\lambda}_i \right)\right]^2
, \label{eq:PCDL1}
\end{align}
where $y$ is the dependent variable, $\mathbf{p}$ and $t$ are the independent variables, and $\bm{\lambda}$ are the parameters of the system. Here, the loss function describes the mean squared error of the residual of the PDE $\mathcal{R}$ (first term), the boundary conditions (second term) and the initial condition (third term), with the number of points sampled given by $N$. One attractive feature of PINNs is the use of automatic differentiation in computing the derivates of the neural network during training~\cite{griewank2008evaluating}, which is done so with standard machine learning libraries~\cite{abadi2016tensorflow, paszke2019pytorch}. As a result, the PDE does not need to be discretized, which enable PINNs to be mesh-free. PINNs thus only require training points to be specified, and a sufficiently low loss implies that the training process results in the output $y$ approximately satisfying the PDE, boundary, and initial conditions, implying a solution to the PDE has been found. Another attractive feature of PINNs is its ability to learn solutions of PDEs for a broad range of parameters $\bm{\lambda}$, enabling parametric solutions of PDEs to be learned at once during training. While the offline training time of PINNs can be large, it only has to train once over all relevant parameters $\bm{\lambda}$ and can be deployed online to make rapid predictions (typically at millisecond timescales). For complex systems that require computationally demanding simulation, many iterations can be efficiently performed using a PINN.

The governing equation for REs is the well known relativistic Fokker-Planck equation, with the inclusion of a large-angle collision operator, where the solution is the electron distribution $f_e$. One challenge with employing a PINN to directly learn the solution to the Fokker-Planck equation is the large variation in the magnitude of $f_e$ for different energies. Specifically, REs typically have energies of several MeV, but their number is much smaller than the total number of electrons. Thus, their contribution to $f_e$ will be orders of magnitude smaller in comparison to the thermal electron population~\cite{guo2017phase,mcdevitt2018relation}. As a result, a PINN trained by directly inputting the residual of the relativistic Fokker-Planck equation into the loss defined by Eq.~(\ref{eq:PCDL1}) will often struggle to resolve the RE tail across a broad range of parameter regimes. An alternative approach in describing RE formation involves the adjoint of the relativistic Fokker-Planck equation~\cite{siegert1951first,karney1986current}, which has been applied in runaway electron formation previously~\cite{liu2016adjoint,liu2016adjointavalanche,zhang2017backward,mcdevitt2023physics,arnaud2024physics,mcdevittpart12025,mcdevittpart22025}. Instead of describing the electron distribution $f_e$, the solution of the adjoint PDE describes the probability of an electron running away, which we will denote by $P$. Since the probability of an electron running away is closely related to the threshold energy for RE formation, this region typically occurs at much lower energies than the characteristic RE energy. Moreover, $P$ is inherently of order unity, thus avoiding the need to resolve the small tail distribution of REs. We proceed with a brief description of the adjoint formulation used in this paper, where an in depth description is provided in Ref.~\cite{mcdevittpart12025}.

\subsection{Steady-state runaway probability function (RPF)}
The adjoint of the steady-state relativistic Fokker-Planck equation is given by
\begin{equation}\label{eq:SSRPF1}
U_p\frac{\partial P}{\partial p} + (1-\xi^2)\left[- \frac{E}{E_c}\frac{1}{p} + \alpha\frac{\xi}{\gamma}\right] \frac{\partial P}{\partial\xi} + \frac{\nu_D}{2}\frac{\partial}{\partial\xi}\left[(1-\xi^2)\frac{\partial P}{\partial\xi}\right] = 0,
\end{equation}
where the independent variables are the electron's momentum $p$ normalized to $m_ec$ and pitch-angle $\xi = p_\Vert/p$, the Lorentz factor is given by $\gamma = \sqrt{1+p^2}$, the characteristic electric field strength is $E/E_c$ with the Connor-Hastie electric field given by $E_c = m_ec/(e\tau_c)$~\cite{connor1975relativistic}, the synchrotron radiation strength is given by $\alpha = \tau_c/\tau_s$ with the relativistic collision timescale $\tau_c = 4\pi\epsilon_0^2m_e^2c^3/(e^4n_e\ln\Lambda)$, free electron density $n_e$, Coulomb logarithm $\ln\Lambda$, characteristic synchrotron radiation damping timescale $\tau_s = 6\pi\epsilon_0m_e^3c^3/(e^4B^2)$, and magnetic field strength $B$. The momentum flux is given by $U_p = -(E/E_c)\xi - C_F - \alpha\gamma p (1 - \xi^2)$, and the collision coefficients are the collisional drag strength $C_F$ and pitch-angle scattering frequency $\nu_D$ (both normalized to $\tau_c$), whose specific form is given by
\begin{equation}\label{eq:SSRPF2}
C_F = 2\left(\frac{c}{v_{Te}}\right)^2\psi(x)\left(\frac{\ln\Lambda_{ee}}{\ln\Lambda_0}\right)\left[1+\frac{1}{\ln\Lambda_{ee}}\sum_k\frac{n_k}{n_e}N_k\left[\sigma^{-1}\ln(1+h_k^\sigma) - \beta^2\right]\right],
\end{equation}
\begin{equation}\label{eq:SSRPF3}
\nu_D = \frac{\gamma}{p^3}\frac{\ln\Lambda_{ei}}{\ln\Lambda_0}Z_{eff}\left\{ 1 + \frac{1}{Z_{eff}\ln\Lambda_{ei}}\sum_k\frac{n_k}{n_e}g_k  + \frac{\ln\Lambda_{ee}}{\ln\Lambda_0} \left[\phi(x) - \psi(x) + \frac{1}{2}\left(\frac{c}{v_{T_e}}\right)^4\right]x^2\right\},
\end{equation}
where $k$ is the atomic species, $x \equiv v/v_{T_e}$ with thermal velocity $v_{T_e}$ and velocity $v = p/\gamma$, $\psi(x)$ is the Chandrasekhar function, the error function is $\phi(x)$, the Coulomb logarithms are $\ln\Lambda_{ee} = \ln\Lambda_0 + (1/\sigma)\ln\{1 + [2(\gamma-1)/p_{T_e}^2]^{\sigma/2}\}$ with thermal momentum $p_{T_e}$, $\ln\Lambda_{ei} = \ln\Lambda_0 + (1/\sigma)\ln[1+(2p/p_{T_e})^\sigma]$, and $\ln\Lambda_0 = 14.9 - 0.5\ln(n_e [10^{20} \mathrm{m}^{-3}]) + \ln(T_e [\mathrm{keV}])$, and $\sigma = 5$~\cite{hesslow2017effect}. Finally, $h_k = p\sqrt{\gamma-1}/I_k$ with mean excitation energy of the ion $I_k$, normalized to the electron rest energy, and $g_k$ is the partially screened contribution [see Eq.~(6) of Ref.~\cite{hesslow2017effect} for an explicit expression].
 
The dependent variable $P$ represents the runaway probability function (RPF), which describes the probability that an electron at an initial momentum space $(p,\xi)$ running away at a later time. The specific definition of the RPF in Eq. (\ref{eq:SSRPF1}) is more subtle, however. Since synchrotron radiation is present, electrons will not run away to infinite energy and instead saturate at some large energy~\cite{decker2016numerical,guo2017phase}. Thus, the RPF here will describe the probability of an electron reaching the high energy boundary $p_{max}$ before saturating at a given energy. As boundary conditions, we will take $P = 1$ at $p_{max}$ when the energy flux $U_p = -(E/E_c)\xi - C_F - \alpha\gamma p (1 - \xi^2)$ is positive, but leave $P$ unconstrained at $p_{max}$ when $U_p < 0$. This latter property follows since when $U_p < 0$ electrons at $p_{max}$ will not immediately run away, and hence we should not force $P$ to one, but these electrons have a finite chance of running away at a later time. As the final boundary condition, we will take $P = 0$ at $p_{min}$, where $p_{min}$ will have a value much lower than the required energy for an electron to run away. For this analysis we will take $p_{min}$ to correspond roughly to the completely screened limit of the force balance between the electric field acceleration and collisional drag [i.e., $p_{min} \approx 1/\sqrt{(E/E_c)-1}$]. An upper limit of $p_{min}$ will also be chosen to correspond to one hundred keV, such that $p_{max} > p_{min}$ remains satisfied.
 
  \begin{figure}
\begin{centering}
\subfigure[]{\includegraphics[width=.5\textwidth]{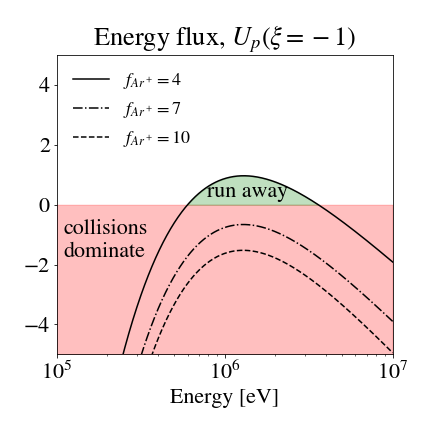}}%
\subfigure[]{\includegraphics[width=.5\textwidth]{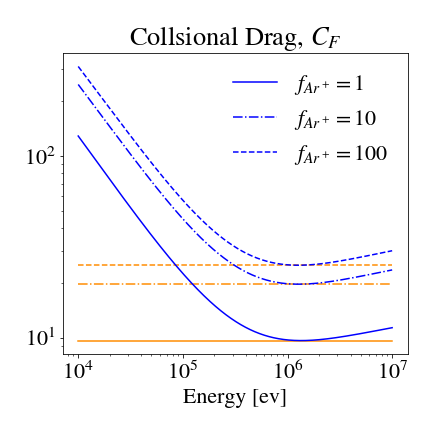}}%
\par\end{centering}
\caption{(a) The energy flux $U_p$ with $\xi = -1$ for various impurity fractions. (b)  The collisional drag $C_F$ as defined in Eq.~(\ref{eq:SSRPF2}) for impurity fractions $f_{Ar^+} = n_{Ar^+}/n_D = [1,10,100]$ and an electric field strength of $E/E_c = 16$. For both panels, a deuterium density of $n_D = 10^{20}$ m$^{-3}$, an impurity species of singly ionized argon, a plasma electron temperature of $T_e = 10$ eV,  and a negligible amount of synchrotron radiation ($\alpha = 0$) was chosen.}
\label{CollDrag}
\end{figure}
 
While $p_{max}$ will typically be chosen to correspond to MeV energies, such a choice here can result in an unconstrained high energy boundary for $P = 1$. Specifically, given the form of the collisional drag $C_F$ in Eq.~(\ref{eq:SSRPF2}), it can be seen that for a given temperature $\ln\Lambda_{ee}$ increases logarithmically in energy, thus leading to $C_F$ increasing at large $p$ and decreasing $U_p$
for a given plasma composition and electric field. Hence, even if the
electric field exceeds collisional drag for modest values of $p$, the
drag will eventually overcome electric field acceleration at
sufficiently high momentum. To illustrate this feature, we plot the
energy dependence of $U_p$ with $\xi= -1$ [see Fig.~\ref{CollDrag}(a)] for various impurity content fractions, and an electric field strength of $E/E_c = 16$. Here, the region where electrons with $\xi = -1$ can run away is given by the shaded green region, where the regions where
collisions dominate, leading to the slowing down of electrons, is
given by the shaded red region. We see the that for energies greater
than approximately 2 MeV, $U_p$ begins to decrease, and as the
impurity fraction increases, the region corresponding to $U_p > 0$
will vanish. As a result, the choice of $p_{max}$ that satisfies
$U_p(\xi=-1) > 0$ will depend on the plasma composition. Another
subtle feature shown in Fig.~\ref{CollDrag}(b) is that the minimum
electric field that satisfies $U_p > 0$ is dependent on the plasma
composition. Specifically, considering an electron with $\xi = -1$,
the energy flux reduces to $U_p = E/E_c - C_F$. Thus, the minimum
$E/E_c$ to satisfy $U_p>0$ will increase with impurity content, as is
shown by the blue and orange curves in Fig.~\ref{CollDrag}(b). For
electric field strengths less than the values corresponding to the
orange curves, the RPF $P$ will vanish everywhere, as no electrons
will be able to run away. When deploying a PINN to learn
Eq.~(\ref{eq:SSRPF1}), we can embed this information directly into the PINN,
which is discussed in the following section.

\subsection{Steady-state RPF PINN}
This section describes the construction of the PINN. The PDE is given by Eq.~(\ref{eq:SSRPF1}), where the momentum in the PINN will be normalized to $p \rightarrow (p-p_{min})/(p_{max}-p_{min})$, and the loss function that will be minimized by the PINN will be taken to have the form:
\begin{equation}
\mathcal{L} = \frac{1}{N_{PDE}} \sum^{N_{PDE}}_i  \left[\left(\frac{1}{\sqrt{E/E_c}}\right)\left( \frac{1}{C_F}\right)\mathcal{R} \left(p_i, \xi_i ; \bm{\lambda}_i \right) \right]^2 
, \label{SSloss}
\end{equation}
where $\mathcal{R}$ is the residual of Eq.~(\ref{eq:SSRPF1}). Here, the $(1/\sqrt{E/E_c})$ and $(1/C_F)$ terms prevent the magnitude of the residual from becoming too large at low energies, or for large electric fields, which can adversely affect the training of the PINN. Specifically, the collisional drag diverges at low momentum $(C_F \propto 1/p^2)$, thus its contribution to $\mathcal{L}$ will result in the PINN primarily focusing on resolving the low energy region. By including the $(1/C_F)$ term in Eq.~(\ref{SSloss}), the loss function will no longer diverge, enabling the PINN to learn a more accurate solution across momentum. Similarly, an analogous process occurs for large electric fields, hence the inclusion of the $(1/\sqrt{E/E_c})$ in Eq.~(\ref{SSloss}). 

While it is straightforward to include boundary conditions ($P = 1$ at $p_{max}$ and $U_p > 0$, $P = 0$ at $p_{min}$) as additional loss terms in Eq.~(\ref{SSloss}), we will directly embed them into the neural network as hard constraints by adding a ``physics layer'' to the output of the PINN. The inclusion of such a physics layer enables customizability of the neural network and can be leveraged to automatically enforce various constraints. In particular, we will construct a physics layer to satisfy the following constraints: (1) the RPF is bounded between zero and unity, (2) a vanishing RPF for electric fields below the threshold for electrons to run away, (3) the RPF vanishes at the low momentum boundary ($p = p_{min}$), and (4) the RPF is unity at the high momentum boundary ($p = p_{max}$) when $U_p > 0$. The constructed physics layer for this analysis is given by:
\begin{subequations}
\label{eq:RPFP1}
\begin{equation}
P^\prime \equiv \Xi_E\left\{\left[\frac{(p-p_{min})\Xi_\xi}{p_{max} - p_{min}}\right] + \left(\frac{p-p_{min}}{p_{max}-p_{min}}\right) \left(\frac{p_{max} - p\ \Xi_\xi}{p_{max}-p_{min}} \right)P_{NN}\right\},
\label{eq:RPFP1a}
\end{equation}
\begin{equation}
P \equiv \tanh\left[\left(\frac{P^{\prime}}{\Delta P}  \right)^2\right],
\label{eq:RPFP1b}
\end{equation}
\end{subequations}
where $P_{NN}$ is the output of the hidden layers of the neural network and the following Heaviside functions are utilized:
\begin{subequations}
\label{eq:RPFP1}
\begin{equation}
\Xi_E \equiv \tanh\left(\frac{E-E^*}{\Delta E_1}\right),
\label{eq:RPFP2a}
\end{equation}
\begin{equation}
\Xi_\xi \equiv \frac{1}{2}\left[1-\tanh\left(\frac{\xi - \xi^*}{\Delta\xi^*}\right)\right].
\label{eq:RPFP2b}
\end{equation}
\end{subequations}
Here, constraint (1) is enforced by Eq.~(\ref{eq:RPFP1b}), and constraint (2) is enforced by the Heaviside function given in Eq.~(\ref{eq:RPFP2a}), where $E^*$ is the minimum electric field to satisfy $U_p > 0$. The low momentum boundary in constraint (3) is satisfied by the ($p-p_{min}$) term in Eq.~(\ref{eq:RPFP1a}), and constraint (4) is satisfied through the term in the square brackets in Eq.~(\ref{eq:RPFP1a}), where a Heaviside function for pitch-angles $\xi$ centered around the critical pitch-angle $\xi^*$ that satisfies $U_p = 0$ is given in Eq.~(\ref{eq:RPFP2b}). 

While constraints (1-4) are satisfied if all the Heaviside functions are taken to be evaluated in the limit where $\Delta P \rightarrow 0$, $\Delta E_1 \rightarrow 0$, and $\Delta\xi^* \rightarrow 0$, such a choice will result in poor PINN training, as the PINN will have to learn a discontinuous function. As a result, we modestly smooth the Heaviside functions with $\Delta P = 0.25$, $\Delta E_1 = 0.25E_c$, and $\Delta\xi^* = 0.15(\tanh[(E - E^*)/\Delta E_2]) + 0.1$ with $\Delta E_2 = 15E_c$. Here, $\Delta\xi^*$ is adaptively chosen depending on $E/E_c$, which enables $\Xi_\xi$ to be modestly smoother for scenarios where $E/E_c \gg E^*/E_c$, but less smooth for $E/E_c \gtrsim E^*/E_c$. We note that the choice of the smoothening will impact the PINN solution near the threshold for RE generation $E/E_c \approx E^*/E_c$; however, the region where the plasma is near the threshold for RE avalanching represents a small contribution towards the broad parameter space we are interested in. A PINN that is tailored to focus specifically in the region where the plasma is near threshold for electrons running away and avalanching has been recently developed~\cite{mcdevittpart12025,mcdevittpart22025}, which could be used in conjunction with the PINNs developed in this work to provide a robust predictor of RE avalanching across a broad range of plasma parameters.

\section{\label{sec:RPR}RPF PINN Results}
This section describes the results of the PINNs deployed to learn the RPF for a broad range of scenarios. The RPF PINN is deployed using the DeepXDE framework~\cite{lu2021deepxde}, along with Tensorflow backend~\cite{abadi2016tensorflow}. During PINN training the ADAM optimizer~\cite{kingma2014adam} is used for the initial iterations, and the L-BFGS-B optimizer~\cite{byrd1995limited} is used for the rest of training. Training points are initialized with a Hammersley distribution, and a residual-based adaptive resampling (RAR) algorithm~\cite{wu2023comprehensive} is deployed, which periodically samples the domain densely to evaluate the residual $\mathcal{R}$. The algorithm then relocates a fraction of training points with a probability function that is proportional to the magnitude of the residual, allowing the PINN to refine the solution near regions with large residuals. Finally, a single Nvidia A100 GPU is used for training the RPF PINN.

\subsection{Verification of a converged RPF PINN solution}
This section demonstrates the characteristics of a converged PINN solution and compares it to ground truth data. Here, we will consider the simplest scenario, where the plasma parameters $\bm\lambda$ are fixed. For the plasma parameters, we will take the electric field strength to be $E/E_c \sim 1000$, the deuterium density to be $n_D = 10^{20}$ m$^{-3}$, the fraction of singly ionized argon density to be $n_{Ar^+}/n_D = 4$, a plasma electron temperature of $T_e = 10$ eV, and no synchrotron radiation $(\alpha = 0)$. The ground truth data is generated from the RunAway Monte Carlo (RAMc) solver, which evolves the guiding center orbits of electrons and includes small-angle collisions, large-angle collisions and synchrotron radiation. For generating the verification data for this section and the remainder of this paper, we have turned off the large-angle collision operator in RAMc, and initialized all particles at the magnetic axis, such that magnetic trapping does not impact the results. Further details of using the Monte Carlo solver to generate the RPF are provided in Ref.~\cite{arnaud2024physics}, with additional information about the RAMc code provided in Ref.~\cite{mcdevitt2019avalanche}.

 \begin{figure}
\begin{centering}
\subfigure[]{\includegraphics[width=.33\textwidth]{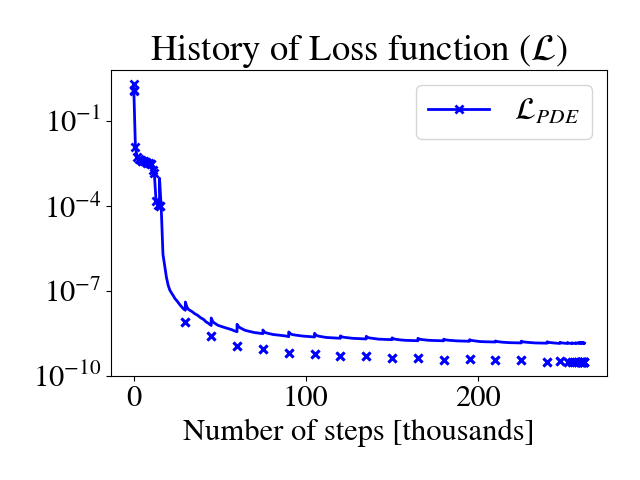}}%
\subfigure[]{\includegraphics[width=.33\textwidth]{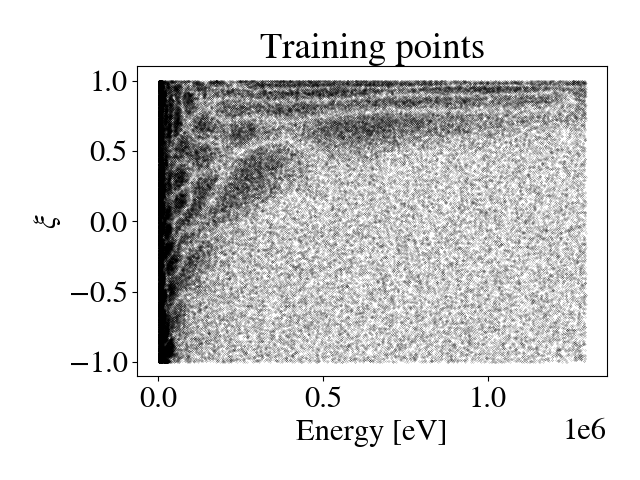}}%
\subfigure[]{\includegraphics[width=.33\textwidth]{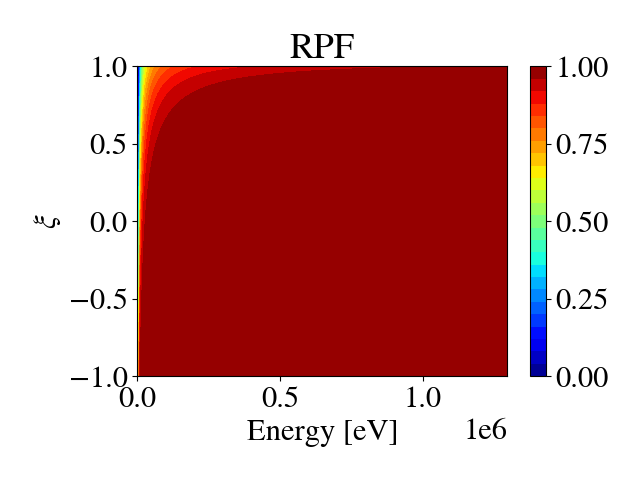}}
\subfigure[]{\includegraphics[width=\textwidth]{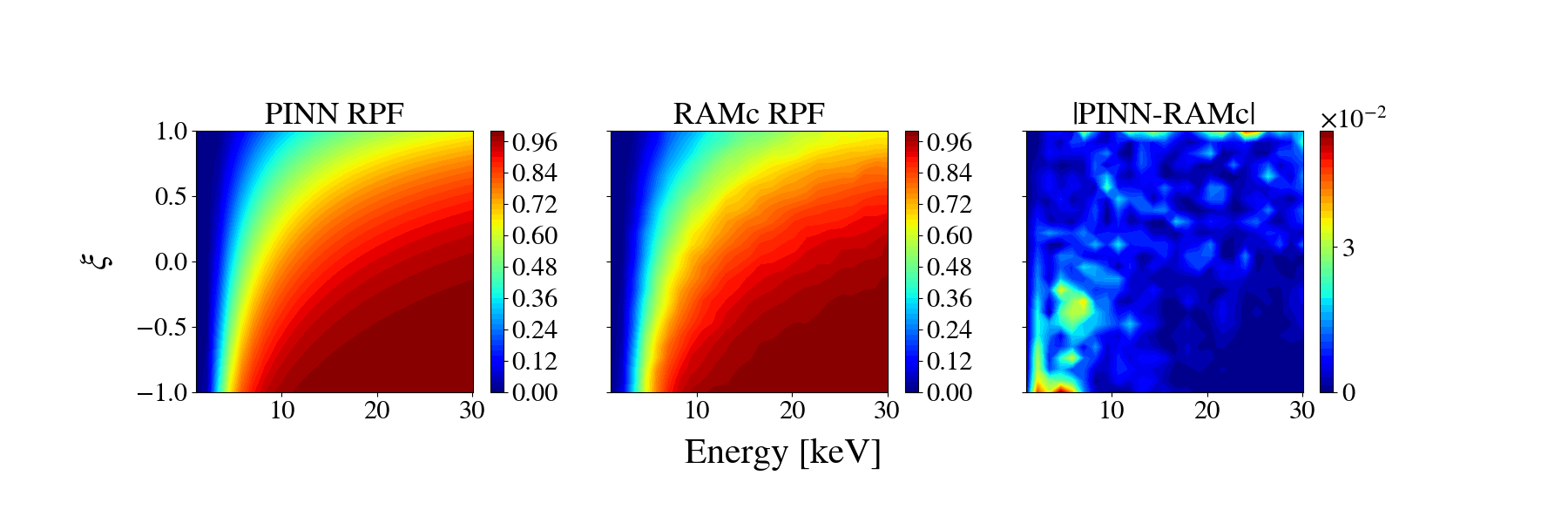}}
\par\end{centering}
\caption{(a) The loss function $\mathcal{L}$ history of the PINN during
  the training period, where the solid curve corresponds to the training point loss, the `x' markers correspond to the test point loss, and the lower test loss is attributed to using a different point distribution and the residual-based adaptive resampling (RAR) algorithm~\cite{wu2023comprehensive}. (b) The final training
  point distribution at the end of the training period. (c) The predicted RPF
  solution from the PINN at the end of the training period. (d)
  A comparison between the predicted RPF solution from the PINN (left column), RAMc (middle
  column), and the corresponding absolute error (right column). The parameters chosen were an electric field strength of $E/E_c \sim 1000$, a deuterium density of $n_D = 10^{20}$ m$^{-3}$, a fraction of singly ionized argon density of $n_{Ar^+}/n_D = 4$, a plasma electron temperature of $T_e = 10$ eV, and no synchrotron radiation $(\alpha = 0)$.}
\label{2DPINNmonteCarloComp}
\end{figure}

We demonstrate a successfully trained PINN and choose the following numerical parameters: one hundred thousand points in the domain of the PDE, an additional number of points along the boundary of the domain, the $\tanh$ activation function for the neurons in the neural network, and four hidden layers with thirty-two neurons per layer. The ADAM optimizer is used for the first 15,000 iterations, and the L-BFGS-B optimizer is used for the rest of training. The results of the PINN after training are shown in Fig.~\ref{2DPINNmonteCarloComp}. The loss function $\mathcal{L}$ is shown in Fig.~\ref{2DPINNmonteCarloComp}(a), where the solid curve corresponds to the loss function associated with the training points and the `x' markers are associated with the test point loss. It can be seen in Fig.~\ref{2DPINNmonteCarloComp}(a) that the PINN successfully learns the solution of the PDE as the loss $\mathcal{L}$ decreases by roughly nine orders of magnitude, where the test points are at modestly lower values. The consistently lower values of the test loss are attributed to the test points being a random uniform distribution in comparison to the Hammersley distribution used for the training points. Additionally, the residual-based adaptive resampling (RAR) algorithm will allocate a portion of training points where the residual of the PDE is largest, which leads to the periodic spikes in the training loss, whereas the test points monotonically decrease and indicates that the RAR algorithm promotes a well-converged solution. In particular, the impact of the RAR algorithm can be seen by looking at the final distribution of training points, as shown in Fig.~\ref{2DPINNmonteCarloComp}(b), where the RAR algorithm emphasizes training points near the separatrix of the RPF [see Fig.~\ref{2DPINNmonteCarloComp}(c)] at a few keV. Once the PINN training converges, the predicted RPF solution is shown in Fig.~\ref{2DPINNmonteCarloComp}(c), where the RPF is predominantly unity for most of the energy range and leads to training points at large energies being relocated to low energies near the separatrix through the RAR algorithm. Finally, a comparison between the predicted RPF solution from the PINN and the ground truth solution computed from RAMc is shown in Fig.~\ref{2DPINNmonteCarloComp}(d), where the comparison is in the range of energies surrounding the separatrix. Here, we see excellent agreement between the PINN and RAMc [see right column of Fig.~\ref{2DPINNmonteCarloComp}(d)], where the remaining discrepancy arises from Monte Carlo noise.

 \begin{figure}
\begin{centering}
\subfigure[]{\includegraphics[width=\textwidth]{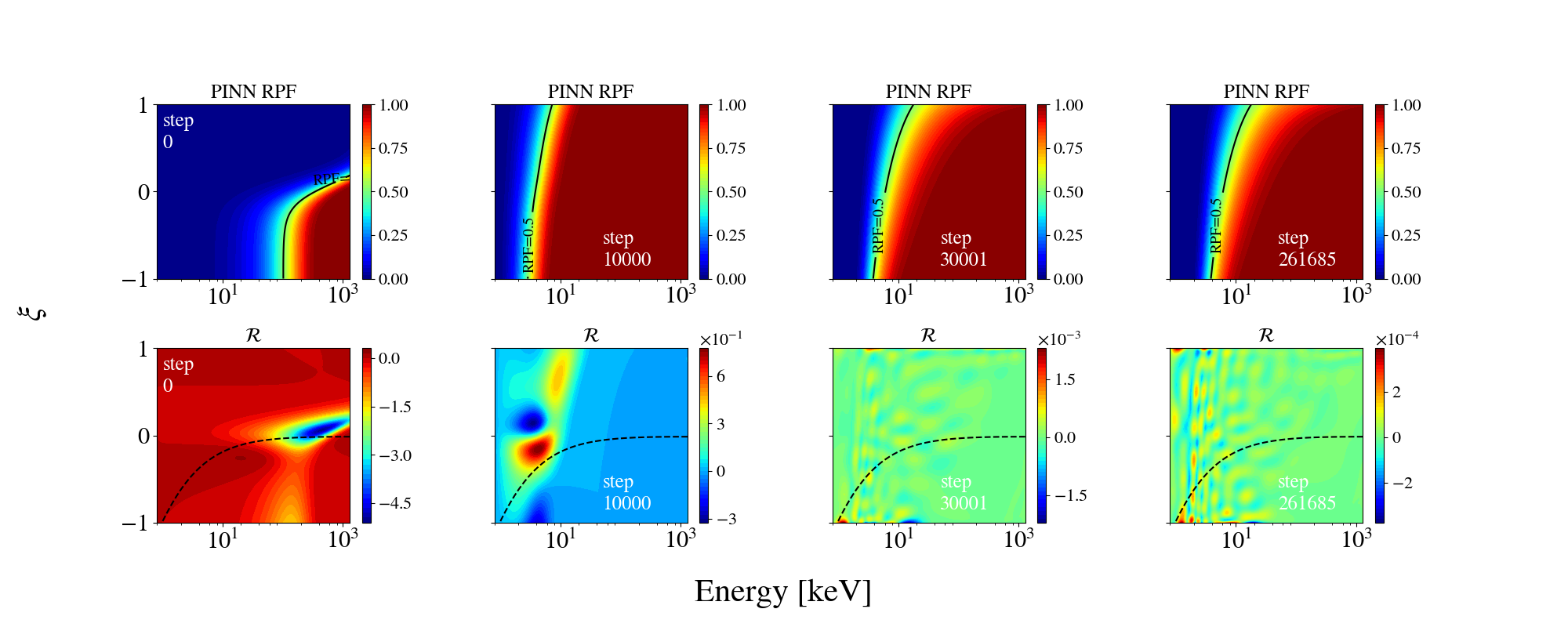}}
\subfigure[]{\includegraphics[width=\textwidth]{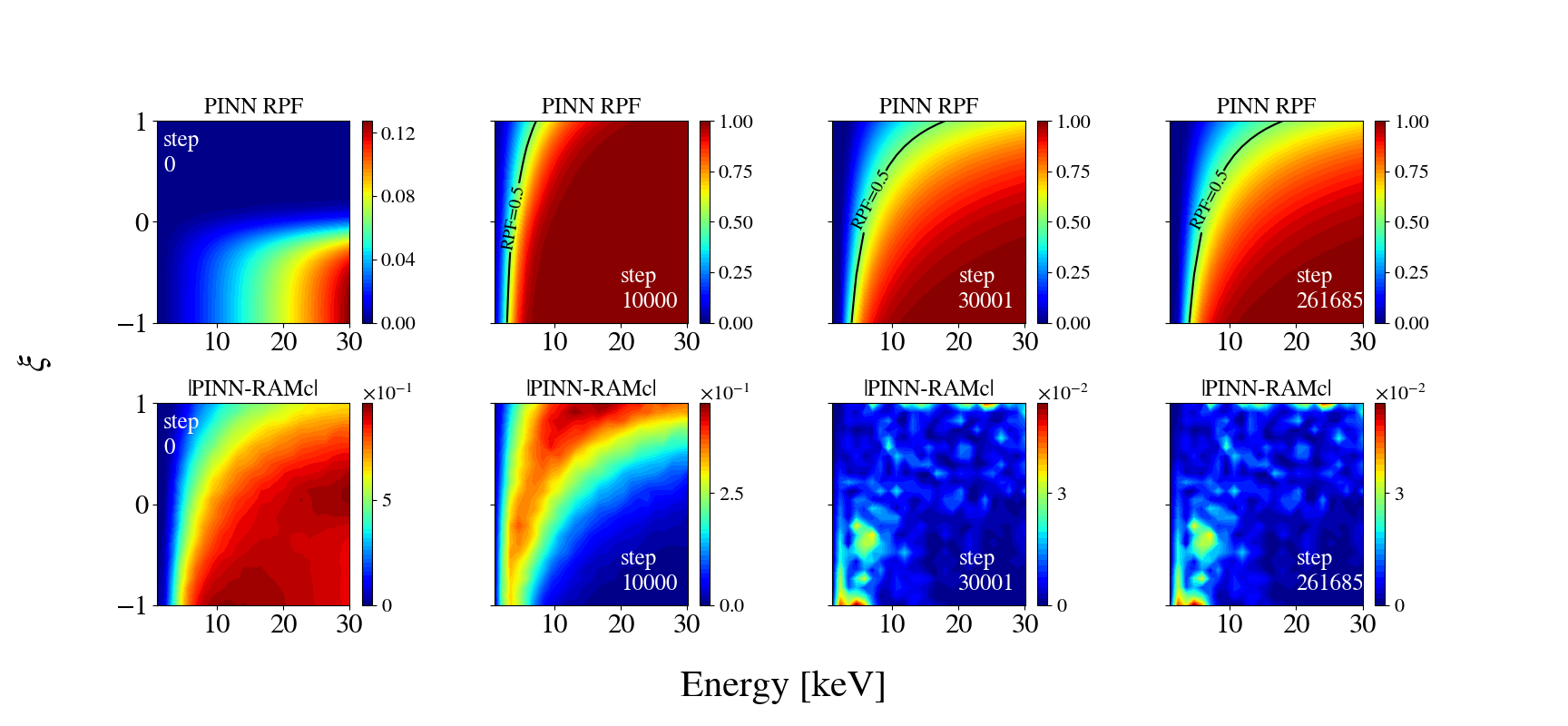}}
\par\end{centering}
\caption{(a) The evolution of the RPF PINN solution (top row) and the residual $\mathcal{R}$ (bottom row) throughout the training period. (b) A comparison between the RPF PINN solution (top row) in the region of energy near the separatrix, and the absolute error between RAMc and the PPF PINN solution (bottom row) throughout the training period. The parameters are the same as that in Fig.~\ref{2DPINNmonteCarloComp}}.
\label{2DPINNmonteCarloCompHistory}
\end{figure}

The evolution of the predicted PINN solution over the training period is shown on the top row of panels in Fig.~\ref{2DPINNmonteCarloCompHistory}(a), and the bottom row of panels in Fig.~\ref{2DPINNmonteCarloCompHistory}(a) show the evolution of the residual $\mathcal{R}$ over the training period, where the dashed black contours correspond to the region (including values of pitch-angle $\xi$ less than that of the dashed-black contours) where $U_p > 0$. Considering the initialization of the PINN solution [see Fig.~\ref{2DPINNmonteCarloCompHistory}(a)], we see that the PINN already identifies the region where $U_p$ is positive near the high energy boundary; however, the residual is large near the region where the $U_p$ transitions from positive to negative. Since the PINN is left unconstrained for pitch-angles $\xi$ at the high momentum boundary $p_{max}$ that correspond to $U_p < 0$, the PINN will not immediately learn the pitch-angle dependence of the separatrix, which can be seen in the second column of Fig.~\ref{2DPINNmonteCarloCompHistory}(a), where the PINN begins to resolve the general location of the separatrix. Subsequently, the PINN learns the pitch-angle dependence of the separatrix [see the third and fourth columns of Fig.~\ref{2DPINNmonteCarloCompHistory}(a)], and the residual is now uniformly low across the entire domain. A comparison of the predicted PINN solution with the RAMc RPF solution is shown in Fig.~\ref{2DPINNmonteCarloCompHistory}(b), where the top rows correspond to the PINN solution at different stages of training and the bottom row is the absolute difference between the true solution and the predicted PINN solution. Here, we see that the absolute error between RAMc and the PINN converges after roughly thirty thousand iterations, where the remaining error is due to the Monte Carlo noise from the RAMc solver.

\subsection{Electric field impact on the RPF}
\label{3D}

\begin{figure}
\begin{centering}
\subfigure[]{\includegraphics[width=0.33\textwidth]{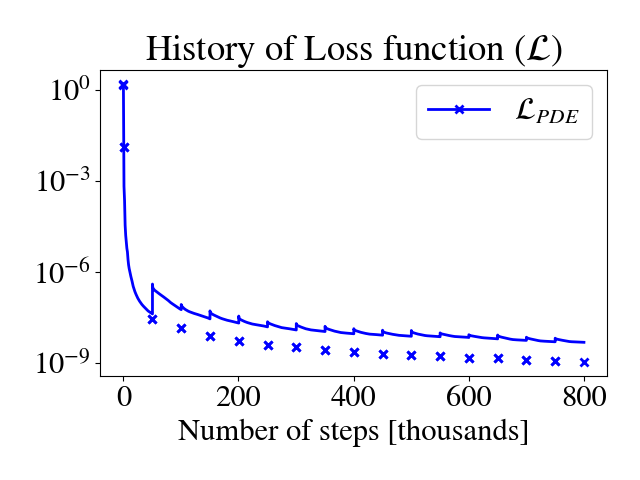}}%
\subfigure[]{\includegraphics[width=0.33\textwidth]{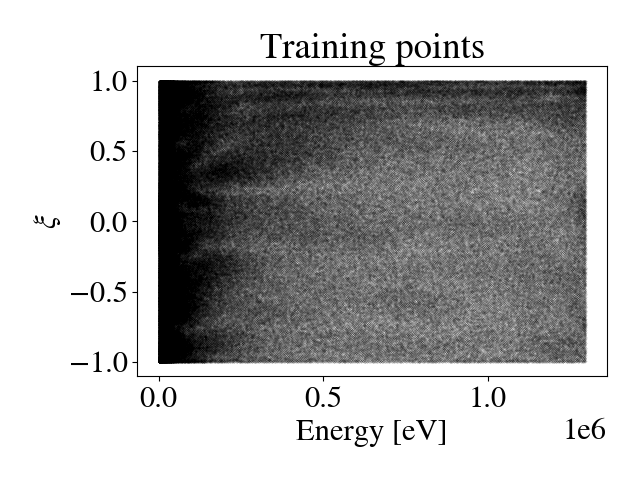}}%
\subfigure[]{\includegraphics[width=0.33\textwidth]{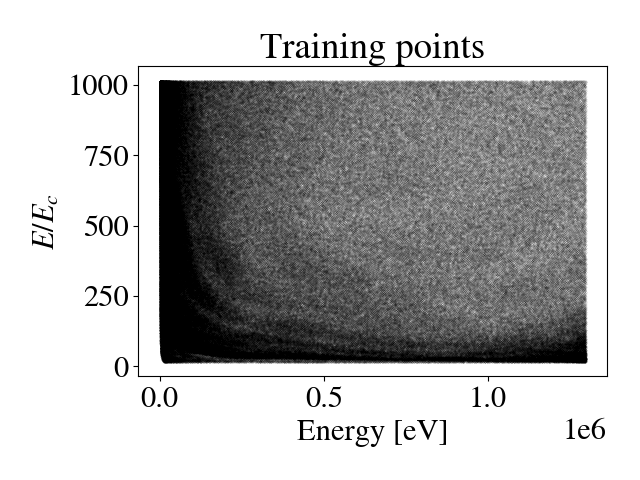}}
\subfigure[]{\includegraphics[width=\textwidth]{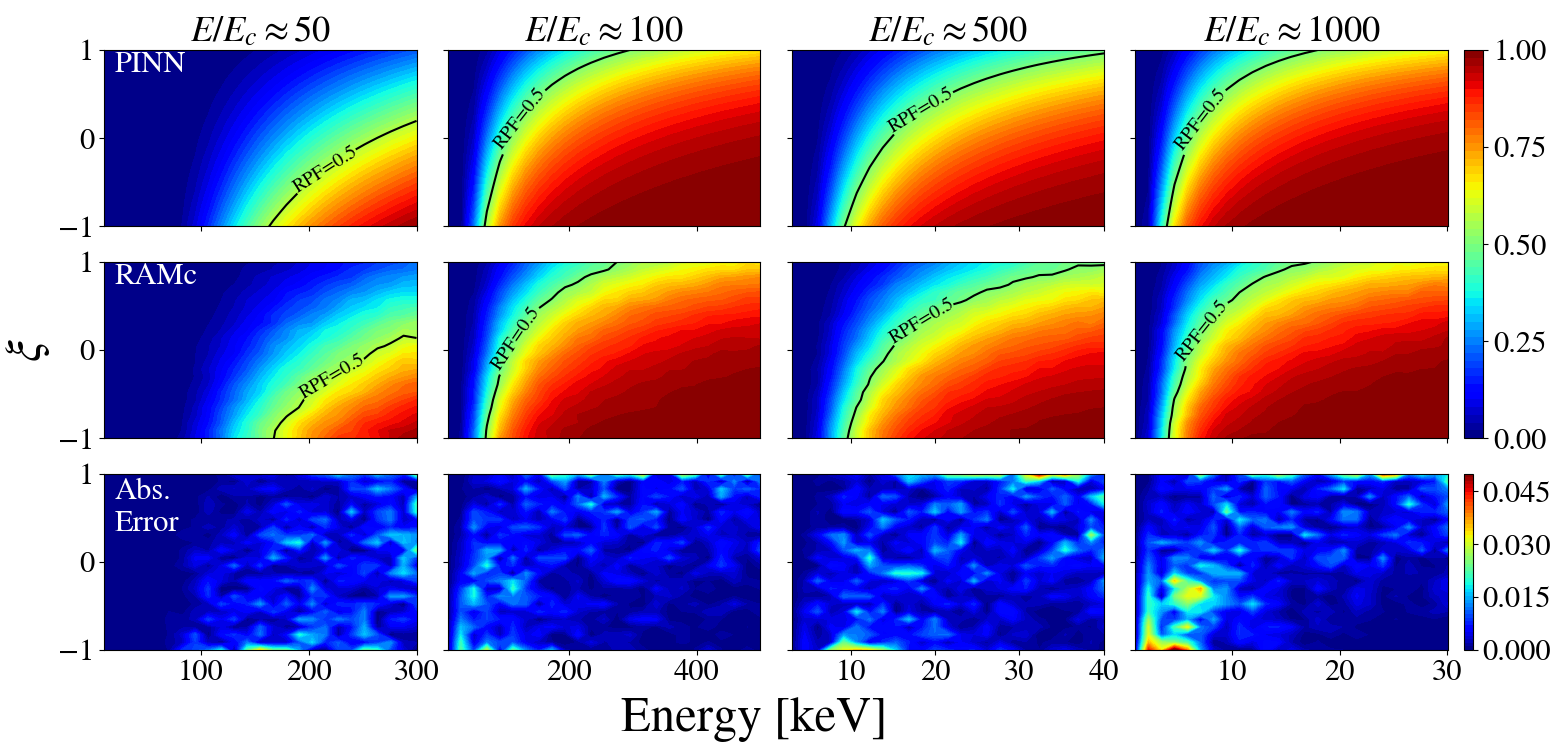}}
\par\end{centering}
\caption{(a) The loss function $\mathcal{L}$ history of the PINN, where the solid curve corresponds to the training loss, and the `x' markers correspond to the test loss. (b,c) The training point distributions at the end of the training period in energy, $\xi$ space [panel (b)] and energy, $E/E_c$ space [panel (c)]. (d) The RPF solution given by the PINN (top row) and the Monte Carlo solver (middle row), and the absolute error between both solutions (bottom row) for varying electric fields strengths. The parameters chosen were a deuterium density of $n_D = 10^{20}$ m$^{-3}$, a fraction of singly charged argon density of 4, a plasma electron temperature of $T_e = 10$ eV, an electric field range of $E/E_c \in [15,1000]$, and no synchrotron radiation $(\alpha = 0)$}.
\label{3DPINNloss}
\end{figure}

In this section we will train the RPF PINN for a range of electric fields ($\lambda = E/E_c$), where we consider a scenario consisting of no synchrotron radiation $(\alpha = 0)$, a deuterium density of $n_D = 10^{20}$ m$^{-3}$, a fraction of singly charged argon density of $n_{Ar^+}/n_D =4$, a plasma electron temperature of $T_e = 10$ eV, and an electric field range of $E/E_c \in [15,1000]$. A fully-connected feed-forward neural network with 64 neurons per layer and 6 hidden layers is used. One million training and test points in the domain, along with a fraction of points along the boundaries of $(p,\xi,E/E_c)$ are sampled. The loss history of the PINN is shown in Fig.~\ref{3DPINNloss}(a), where it can be seen that the test loss (blue `x' markers) reaches approximately $10^{-9}$, indicating that the PINN successfully learns the PDE. We see that the final distribution of training points in energy and $\xi$ [see Fig.~\ref{3DPINNloss}(b)] remains concentrated at relatively low energies, similar to the PINN in the previous section; however, the inclusion of a varying electric field results in an additional emphasis at low electric fields and larger energies [see Fig.~\ref{3DPINNloss}(c)], which corresponds to the RPF approaching marginality.

The performance of the PINN is shown in Fig.~\ref{3DPINNloss}(d), where the PINN predictions of the RPF (top row) are compared against the Monte Carlo predictions (middle row), alongside the absolute error (bottom row). Excellent agreement is observed between the PINN and Monte Carlo solutions, where the remaining difference is due to the Monte Carlo noise. An immediate trend present in the RPF solutions are the impact of an increasing electric field, which decreases the threshold energy and pitch-angle (black $P = 0.5$ contour) for electrons to run away, where at $\xi = -1$ the threshold RE energy decreases from roughly a hundred keV to a few keV as the electric field increases from $E/E_c \approx 50$ to $E/E_c \approx 1000$.  We note that the domain of the PINN shown in this section can be increased in capability to also learn the entire parametric solution of the PDE. The relevant parameters would then be $\bm{\lambda} = (E/E_c, n_D, n_k^{(i)}, T_e, B)$, where the density $n_k^{(i)}$ is for each species $k$ and charge state $i$, which consequently increases the dimensionality of the parameter space significantly. Considering neon and argon as the typical impurity species found in disruptions, a PINN would have to learn eleven or nineteen charge states for either neon or argon, thus requiring a significant amount of training points. As a result, the required number of training points will surpass the memory available on a single Nvidia A100 GPU (80 GB) with the current neural network architecture (6 hidden layers and 64 neurons per hidden layer). Therefore, the PINN would need to be parallelized to train on multiple GPUs. While distributed frameworks for multiple GPU usage is readily available, where Horovod~\cite{sergeev2018horovod} is a commonly used framework implemented in DeepXDE, in this work we will overcome this challenge by constraining the PINN to train across physically consistent parameters, as described in Sec. \ref{CRav} below.

\subsection{Embedding a steady-state power balance model to reduce dimensionality}
\label{CRav}
While having a PINN that can predict the RPF solution for a wide range of physics parameters can be attractive for optimization scenarios that conduct a many queries analysis, the physics parameters $\bm{\lambda}$ are often correlated in the context of tokamak disruptions. Specifically, the average charge state of an impurity $\bar{Z}_k(T_e)$ depends on the temperature. Thus, for a given plasma composition and temperature, the full parameter space given in the previous section reduces to $\boldsymbol{\lambda} = (E/E_c, n_D, n_k, T_e, B)$, where the reduction in parameter space is $n_k^{(i)} \rightarrow n_k$. A further reduction of parameter space can be achieved by enforcing power balance to evaluate $T_e$, which will be evaluated from a given plasma composition and electric field $(E/E_c,n_D,n_k)$, thus removing an additional input into the PINN. Furthermore, noting that electric field can be evaluated from Ohm's law $E = \eta \left( T_e\right) j$, this then allows us to remove the electric field $E$ as an input in place of the current density $j$. As a result, the parameter space will reduce to $\boldsymbol{\lambda} = (n_D, n_k, j,B)$. If we further develop a PINN that is specific to a specific tokamak with a given magnetic field (for example, $B = 5.3$ tesla for ITER), this further reduces the parameter space to $\boldsymbol{\lambda} = (n_D, n_k, j)$.

\begin{figure}
\begin{centering}
\subfigure[]{\includegraphics[width=0.33\textwidth]{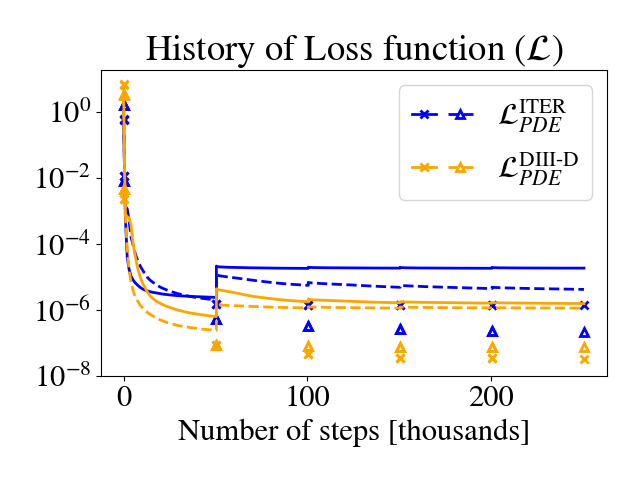}}%
\subfigure[]{\includegraphics[width=0.33\textwidth]{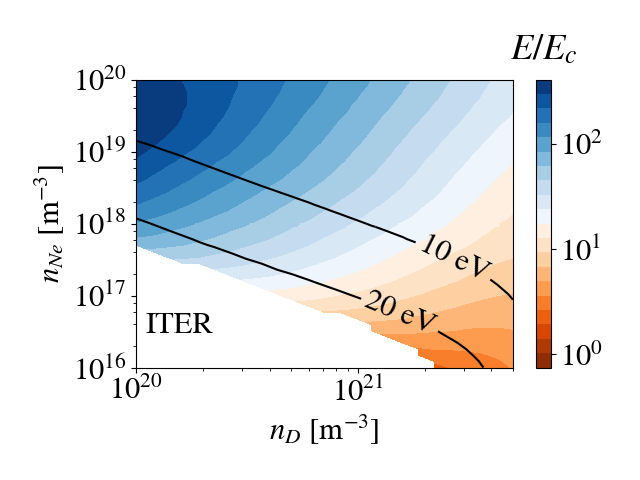}}
\subfigure[]{\includegraphics[width=0.33\textwidth]{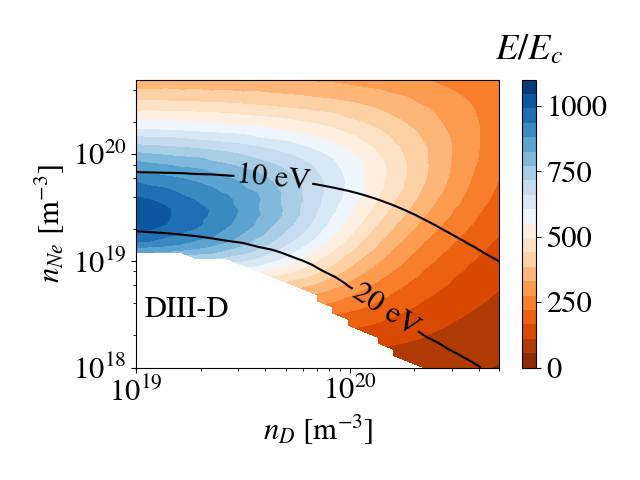}}
\subfigure[]{\includegraphics[width=\textwidth]{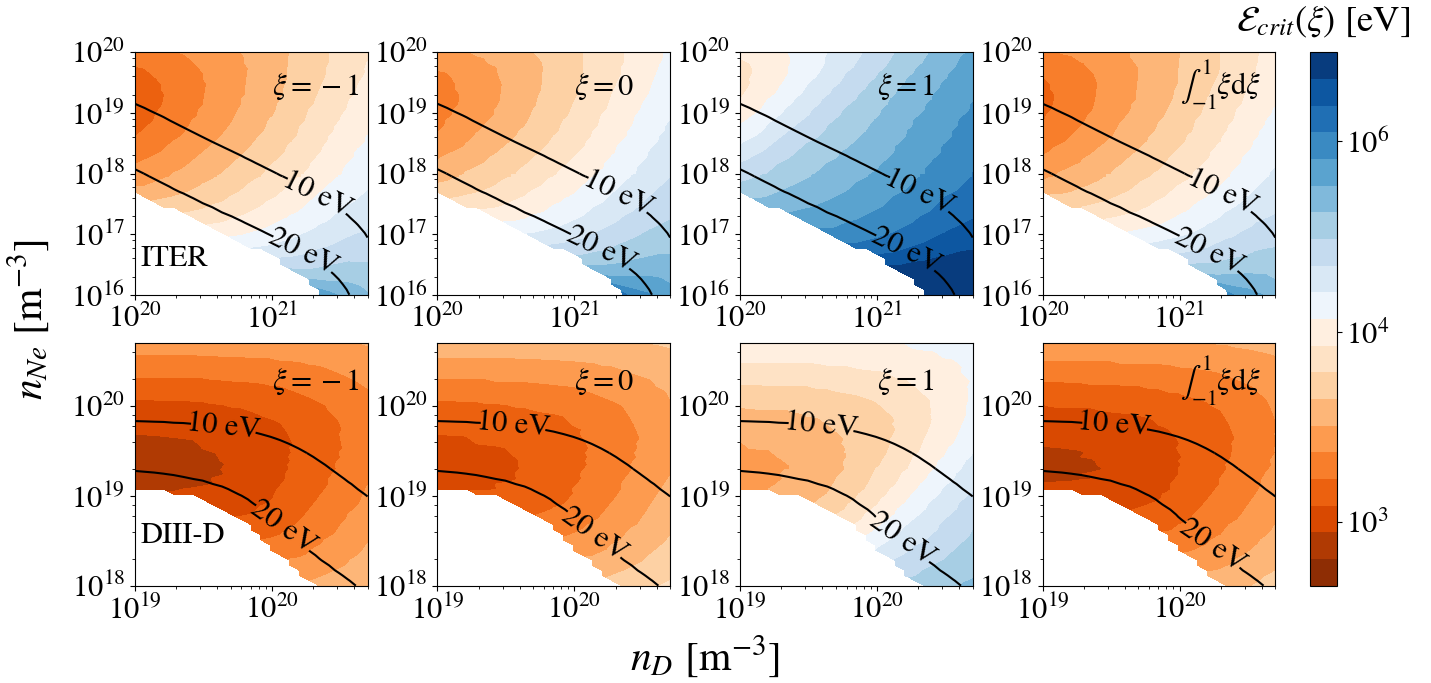}}
\par\end{centering}
\caption{(a) The loss function $\mathcal{L}$ history for the PINNs corresponding to the ITER (orange curves) and DIII-D (blue curves) scenarios, where the curves correspond to the training loss, and the triangle and `x' makers correspond to the test loss. The dashed curve with triangle markers, and the solid curve with `x' markers correspond to the specific subdomain of neon density. (b-c) The electric field strength for the ITER (b) and DIII-D (c) scenario, where the black contours correspond to equilibrium temperatures of ten and twenty eV that is set by a plasma composition and current density. For the ITER scenario $j=1$ MA/m$^2$, and $j=2$ MA/m$^2$ for the DIII-D scenario. (d) The energy in which the RPF is one half ($\mathcal{E}_{crit}$) in eV at various values of pitch-angle $\xi$.}
\label{CRavPINN}
\end{figure}

A PINN tailored to tokamak disruptions is thus demonstrated by deploying a PINN that has an embedded steady-state power balance model with $\bm{\lambda} = \left( n_D, n_{Ne}, j\right)$, and neon is chosen as the impurity element.  The steady-state power balance model consists of a zero dimensional power balance between the Ohmic heating and radiative losses, where the radiative losses are evaluated from data generated from a collisional radiative model (see Refs.~\cite{mcdevitt2023constraint,arnaud2024impact} and the references therein for further details). We will deploy two separate PINNs to learn tokamak operating scenarios representative of the DIII-D tokamak ($B = 2.2$ tesla) and the ITER tokamak ($B = 5.3$ tesla), where the chosen plasma composition domains corresponding to the DIII-D scenario will consist of $n_D \in [10^{19}, 5\times10^{20}]$ m$^{-3}$, $n_{Ne} \in [10^{18}, 5\times10^{20}]$ m$^{-3}$, and the domain corresponding to the ITER scenario will be $n_D \in [10^{20}, 5\times10^{21}]$ m$^{-3}$, $n_{Ne} \in [10^{16}, 10^{20}]$ m$^{-3}$. For both scenarios the current density range will be chosen to be a modest range of $j \in [0.75, 2.25]$ MA/m$^2$. We note that since the range of parameters chosen encompasses a broad range of relevant disruption scenarios, we will be able to obtain better converged solutions by learning batches of the subdomain space. Specifically, we shall decompose the impurity density range to the following subranges: $n_{Ne} \in [10^{16},10^{18}]$ m$^{-3}$ and $n_{Ne} \in [9 \times 10^{17},10^{20}]$ m$^{-3}$ for the ITER scenario, and $n_{Ne} \in [10^{18},3\times10^{19}]$ m$^{-3}$ and $n_{Ne} \in [2 \times 10^{19},5 \times 10^{20}]$ m$^{-3}$ for the DIII-D scenario, thus four total PINNs. Two million training points with a fully connected neural network containing 32 neurons and 4 hidden layers will be used for each PINN.

The loss histories of the four PINNs are shown in Fig.~\ref{CRavPINN}(a), where the orange and blue curves correspond to the DIII-D and ITER scenarios, respectively, the solid curves correspond to $n_{Ne} \in [10^{16}, 10^{18}]$ m$^{-3}$ (blue) for the ITER scenario and $n_{Ne} \in [10^{18}, 3\times 10^{19}]$ m$^{-3}$ (orange) for the DIII-D scenario, and the dashed curves correspond to $n_{Ne} \in [9\times 10^{17}, 10^{20}]$ m$^{-3}$ (blue) for the ITER scenario and $n_{Ne} \in [2\times 10^{19},5\times 10^{20}]$ m$^{-3}$ (orange) for the DIII-D scenario. Here, we see that the PINN is able to rapidly learn the RPF solution in as few as one hundred thousand steps, where the test loss for all PINNs saturates at roughly between values of $10^{-6} \sim 10^{-8}$. Both the training and test losses are seen to be modestly lower for the DIII-D scenario, which arises from the $1/\sqrt{E/E_c}$ term present in Eq.~(\ref{SSloss}). Noting that $E_c \propto n_e$, the DIII-D scenario will contain larger $E/E_c$ in comparison to the ITER scenario. This is shown in Figs.~\ref{CRavPINN}(b) and (c) for the ITER and DIII-D scenario, respectively, where the maximum $E/E_c$ is nearly an order of magnitude larger than the ITER scenario and both scenarios have $E/E_c$ peaked at low $n_D$ and $n_{Ne}/n_{D} \approx 1$. The black contours represent equilibrium temperatures of ten and twenty eV, which approximately correspond to the targeted thermal quench times for ITER. The white region represents scenarios where the plasma reheats to temperatures greater than one hundred eV and results in $E/E^* < 1$, thus leading to the RPF vanishing.
 
The impact of plasma composition and current density on the RPF can be characterized by the energy and pitch-angle where the RPF is one half, thus we define the critical energy and critical pitch-angle ($\mathcal{E}_{crit}, \xi_{crit}$) to be the $P=0.5$ contour. The resulting calculation is shown in Fig.~\ref{CRavPINN}(d), where the rough correlation $\mathcal{E}_{crit} \propto (E/E_c)^{-1}$ can be seen. As $\xi_{crit}$ increases [compare Fig.~\ref{CRavPINN}(d) from the left column to the second to last column on the right], $\mathcal{E}_{crit}$ increases by nearly an order of magnitude, thus electrons with large pitch-angles will have a lower chance of running away at a given energy in comparison to electrons at moderate to small pitch-angles [see RPF solutions in the previous sections].

\section{\label{sec:SSRES}steady-state runaway electron surrogate}
 
With the various PINNs deployed in the previous section to learn the RPF for a broad range of plasma parameters, we turn to utilizing the PINN as a surrogate for predicting the RE avalanche. The rate of REs generated from an arbitrary source of electrons is given by~\cite{liu2016adjointavalanche}
\begin{equation}\label{dnREdt}
\frac{\mathrm{d}n_{RE}}{\mathrm{d}t} \equiv \int \mathrm{d^3p}S(p,\xi,\boldsymbol{\lambda})P(p,\xi,\boldsymbol{\lambda}),
\end{equation}
where $S(p,\xi,\boldsymbol{\lambda})$ represents a source with relevant physics parameters for the scenario of interest $\boldsymbol{\lambda}$, and the volume element is $\mathrm{d^3p} \equiv 2\pi p^2\mathrm{d}p\mathrm{d}\xi$. We note that while this paper focuses on a source of electrons $S(p,\xi,\bm{\lambda})$ generated by large-angle Coulomb collisions, $S(p,\xi,\bm{\lambda})$ can represent an arbitrary source, if one interested in RE generation from $\beta$ decay of tritium, or Compton scattering from an activated wall~\cite{rosenbluth1997runaway,martin2017formation}. We leave the development of a surrogate describing the aforementioned mechanisms as future work.

Considering the scenario where REs are already present in the plasma, large-angle collisions can occur between an existing RE (primary electron) with momentum and pitch-angle ($p^\prime,\xi^\prime$) and an initially cold thermal electron, which after the large-angle collision has a momentum and pitch-angle of ($p,\xi$) (secondary electron). The source of secondary electrons $S(p,\xi)$ after a large-angle collision is given by:
\begin{equation}
S(p,\xi) = \frac{n_{total}cr_e^2}{2\pi} \int \mathrm{d^3p^\prime}\frac{v^\prime}{p^2}\frac{d\sigma_M(p^\prime,p)}{dp}\Pi(p^\prime,\xi^\prime,p;\xi) f_e(p^\prime,\xi^\prime) 
\label{Moller}
\end{equation}
where the primary electron's velocity is normalized to $v^\prime \rightarrow v^\prime/c$, the M\o ller cross section~\cite{moller1932theorie} is normalized to $\sigma_M \rightarrow \sigma_M/r_e^2$, a constraint on the secondary electron's pitch-angle is given by $\Pi(p^\prime,\xi^\prime,p;\xi)$, the total electron density is given by $n_{total} \equiv n_e + n_{bound}$, including free $n_e$ and bound $n_b$ electrons, the classical electron radius is given by $r_e = e^2/(4\pi\epsilon_0m_ec^2)$, the primary electron distribution is $f_e(p^\prime,\xi^\prime)$, and $\mathrm{d^3p^\prime} \equiv 2\pi p^\prime\mathrm{d}p^\prime\mathrm{d}\xi^\prime$ is the volume element. The specific form of the M\o ller cross section used is given by
\begin{equation}
\frac{\mathrm{d}\sigma_M(\gamma^\prime,\gamma)}{\mathrm{d}p} = \frac{2\pi v\gamma^{\prime^2}}{(\gamma^\prime-1)^3(\gamma^\prime+1)}\left[x^2 - 3x + \left(\frac{\gamma^\prime-1}{\gamma^\prime}\right)^2(1+x)\right]
, \label{eq:Moller}
\end{equation}
where we have introduced the Lorentz factor of the primary electron $\gamma^{\prime} \equiv \sqrt{1+p^{\prime^2}}$ for convenience and $x \equiv (\gamma^\prime-1)/\{(\gamma-1)[1-(\gamma-1)/(\gamma^\prime-1)]\}$. The secondary pitch-angle dependence is set by~\cite{boozer2015theory}
\begin{equation}
\Pi(\gamma^\prime,\xi^\prime,\gamma;\xi) = \frac{1}{\pi}\frac{1}{\sqrt{\xi_2^2 - (\xi - \xi_1)^2}},
\end{equation}
\begin{equation}
\xi_1(\gamma,\gamma^\prime,\xi^\prime) = \xi^\prime \sqrt{\frac{(\gamma^\prime+1)(\gamma-1)}{(\gamma^\prime-1)(\gamma+1)}},
\end{equation}
\begin{equation}
\xi_2(\gamma,\gamma^\prime,\xi^\prime) = \sqrt{\frac{2(\gamma^\prime-\gamma)}{(\gamma^\prime-1)(\gamma+1)}(1-\xi^{\prime^2})},
\end{equation}
for secondary pitch-angles that satisfies $|\xi - \xi_1|\leq \xi_2$, otherwise $\Pi(\gamma^\prime,\xi^\prime,\gamma;\xi) = 0$. We note that the secondary source $S(p,\xi)$ requires knowledge of the primary electron distribution $f_e(p^\prime,\xi^\prime)$, which is not explicitly given by the adjoint PDE described in Eq.~(\ref{eq:SSRPF1}), thus requiring a closure.

\subsection{Rosenbluth-Putvinski source}\label{RP}
The simplest closure, introduced in Ref.~\cite{rosenbluth1997theory}, is to assume an existing runaway electron population with asymptotically large energies and completely aligned with the magnetic field $(\xi^\prime = -1)$. Taking $p^\prime \rightarrow \infty$ and $\xi^\prime = -1$ results in $\mathrm{d}\sigma_M(\infty,\gamma)/\mathrm{d}p \rightarrow 2\pi v/(\gamma-1)^2$, thus $S(p,\xi)$ reduces to
\begin{equation}
S_{RP} (p, \xi)  = n_{total} n_{RE} c r^2_e \frac{v}{\gamma^2-1} \frac{1}{\left( \gamma-1 \right)^2} \delta \left( \xi - \xi_1 \right),
\end{equation}
where $\Pi(\gamma,\infty,-1) \rightarrow \delta(\xi - \xi_1)$, $\xi_1(\gamma,\infty,-1) = -\sqrt{(\gamma-1)/(\gamma+1)}$, and $n_{RE} \equiv \int \mathrm{d^3p^\prime}f_e(p^\prime,\xi^\prime)$ $\approx \int 2\pi p^{\prime^2}\mathrm{d}p^\prime\mathrm{d}\xi^\prime f_e(\infty,-1)$. Equation (\ref{dnREdt}) then takes the form:
\begin{equation}
\left. \frac{d n_{RE}}{dt} \right|_{av} = 2\pi n_{total} n_{RE} c r^2_e \int \mathrm{d}p \frac{v}{\left( \gamma - 1 \right)^2} P \left( p, \xi_1 \right),
\end{equation}
where the exponential `avalanche' growth rate normalized to the relativistic collision time $\tau_c$, after some algebra, can be defined to be
\begin{equation}
\gamma_{av}^{RP}\tau_c \equiv \frac{\left. \frac{d n_{RE}}{dt} \right|_{av}}{n_{RE}}\tau_c = \frac{n_{total}}{2n_e\ln\Lambda}\int \mathrm{d}p \frac{v}{\left( \gamma - 1 \right)^2} P \left( p, \xi_1 \right).
\label{gammaAvEq}
\end{equation}
The avalanche growth rate using the Rosenbluth-Putvisnki source $S_{RP}$ has been shown to be accurate for scenarios modestly above marginality~\cite{mcdevitt2018relation,arnaud2024physics}; however, the assumption of primary electrons having asymptotically large energies is likely to break down for scenarios near marginality, as electrons will not be accelerated to large energies, and RE avalanching will be over predicted in this regime, thus underestimating the threshold for avalanching. Moreover, for plasma scenarios containing significantly large electric fields [recall Figs.~\ref{CRavPINN}(b,c)], the threshold energy for RE generation will decrease significantly [see Fig.~\ref{3DPINNloss}(d)]. Noting that the M\o ller cross section is peaked at lower primary electron energies [see Fig. 2 of Ref.~\cite{chiu1998fokker}], modestly relativistic primary electrons will ultimately lead to a larger contribution of the source of secondary electrons $S(p,\xi)$ in comparison to the Rosenbluth-Putvinski assumption of asymptotically large energies. As a result, for scenarios containing large electric fields, the avalanche growth rate $\gamma_{av}$ will be under predicted by $S_{RP}(p,\xi)$. Thus, a higher fidelity closure of the primary electron distribution is required for a more robust prediction of the avalanche growth rate.

\subsection{Chiu-Harvey source}\label{CH}
An improvement can be made over the $p^\prime \rightarrow \infty$ assumption previously made in the Rosenbluth-Putvinski source, where primary electrons will still be assumed to have $\xi^\prime = -1$~\cite{chiu1998fokker,harvey2000runaway}, but the energy distribution of REs will be incorporated. The source of secondary electrons given in Eq.~(\ref{Moller}) is then
\begin{equation}
S(p,\xi) = \frac{n_{total}cr_e^2}{p^2}\int \mathrm{d}p^\prime p^{\prime^2}v^\prime\frac{\mathrm{d}\sigma_M(\gamma^\prime,\gamma)}{\mathrm{d}p}\delta(\xi-\xi_1)F(p^\prime),
\label{Eq:Schiu1}
\end{equation}
where the pitch-angle integrated distribution is defined as $F(p^\prime) \equiv \int\mathrm{d}\xi^\prime f_e(p^\prime,\xi^\prime)$, and the secondary pitch-angle is now $\xi_1(\gamma,\gamma^\prime,-1) = -\sqrt{[(\gamma^\prime+1)(\gamma-1)]/[(\gamma^\prime-1)(\gamma+1)]}$. The integral over $p^\prime$ can be done by noting that $\delta(\xi-\xi_1) = \delta(p^\prime-p^\prime_0)/|\partial(\xi-\xi_1)/\partial p^\prime|_{p^\prime=p^\prime_0}$, where $p^\prime_0$ is the root of $\xi-\xi_1 = 0$, which, after some algebra, is
\begin{equation}
p^\prime_0 = \frac{2p\xi}{1+\xi^2-\gamma(1-\xi^2)}
, \label{eq:p0}
\end{equation}
where we note that $p_0^\prime$ is positive definite from the kinematic constraints given by $\xi_1$ and the fact that the maximum energy a secondary electron can partake is half of the primary electron's energy,
 and $|\partial(\xi-\xi_1)/\partial p^\prime |_{p^\prime=p^\prime_0} = |\xi_1|/(p^\prime_0\gamma^\prime_0)$. Equation~(\ref{Eq:Schiu1}) then reduces to
\begin{equation}
S_{CH}(p,\xi) = \frac{n_{RE}n_{total}cr_e^2}{p^2}\frac{\mathrm{d}\sigma_M(p^\prime_0,p)}{\mathrm{d}p}\frac{p^{\prime^4}_0F(p^\prime_0)}{|\xi_1|},
\end{equation}
where $F(p^\prime_0)$ is normalized to satisfy $n_{RE} = \int\mathrm{d^3p^\prime}f_e(p^\prime,\xi^\prime)$. After some algebra, the avalanche growth rate is then
\begin{equation}
\gamma_{av}^{CH}\tau_c = \frac{n_{total}}{2n_e\ln\Lambda}\int\mathrm{d}p\mathrm{d}\xi\frac{\mathrm{d}\sigma_M(p^\prime_0,p)}{\mathrm{d}p}\frac{F(p^\prime_0)p^{\prime^4}_0}{|\xi_1|}P(p,\xi).
\end{equation}
While the primary electron energy distribution $F(p^\prime)$ is not directly predicted by the solution to the adjoint problem, an approximate form can be used by noting that the steady state RE energy distribution follows a decaying exponential $2\pi p^{\prime^2} F(p^\prime) \propto \exp[-(\gamma^\prime-1)m_ec^2/T_{av}]$, with an `avalanche temperature' given by $T_{av} \equiv m_ec^2(E/E_c-1)/(\gamma_{av}\tau_c)$~\cite{rosenbluth1997theory}. By using $\gamma_{av}^{RP}\tau_c$ given in Eq. (\ref{gammaAvEq}) to approximate $T_{av}$, this allows an initial form for $F \left( p^\prime\right)$ to be identified, and hence an initial estimate of $\gamma_{av}^{CH}\tau_c$ to be computed. By continuing to iterate, a converged value of $\gamma_{av}^{CH}\tau_c$ can be obtained.

\begin{figure}
\begin{centering}
\subfigure[]{\includegraphics[width=0.5\textwidth]{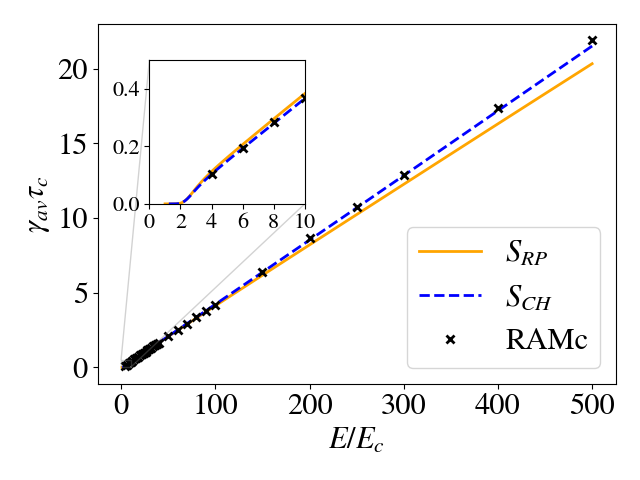}}%
\subfigure[]{\includegraphics[width=0.5\textwidth]{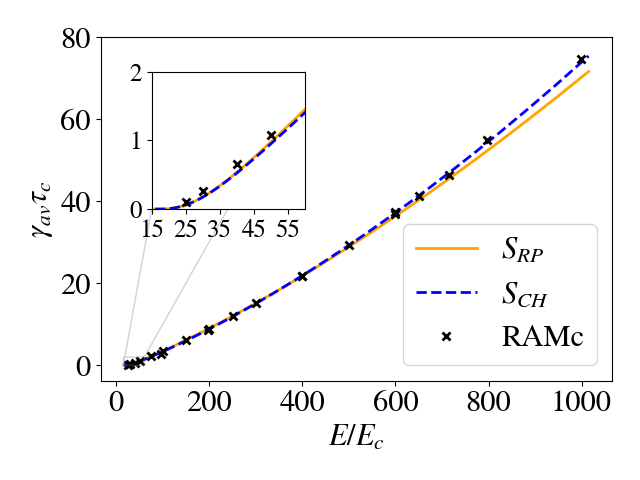}}%
\par\end{centering}
\caption{(a-b) An avalanche growth rate comparison between kinetic Monte Carlo simulation (black `x' markers) and the PINN predictions using the Rosenbluth-Putvisnki source (solid orange curve) and the Chiu-Harvey source (dashed blue curve), where (a) utilizes a PINN from Ref.~\cite{arnaud2024physics} and (b) utilizes the PINN shown in Sec.~\ref{3D}.  The parameters chosen for (a) were a fully ionized plasma with an effective charge of $Z_{eff} = 5$, a synchrotron radiation strength of $\alpha = 0.1$, and an electric field range of $E/E_c \in [1, 500]$, and the parameters chosen for (b) were a deuterium density of $n_D = 10^{20}$ m$^{-3}$, a fraction of singly ionized argon density of $n_{Ar^+}/n_D = 4$, a plasma electron temperature of $T_e = 10$ eV, an electric field range of $E/E_c \in [15, 1000]$, and negligible synchrotron radiation $(\alpha = 0)$.}
\label{3DPINNgammaAv}
\end{figure}

\subsection{RE avalanche comparison between sources}
The performance of the two sources introduced in the previous subsections are compared by utilizing the PINN introduced in Sec.~\ref{3D}, where a range of electric fields are chosen as the plasma parameters to be learned by the PINN. In addition, we extend the PINN developed in Ref.~\cite{arnaud2024physics} to utilize the same loss function in Eq.~(\ref{SSloss}) and physics layer in Eqs.~(\ref{eq:RPFP1a}-\ref{eq:RPFP2b}), where we consider an electric field range consisting of $E/E_c \in [1, 500]$, a synchrotron radiation strength of $\alpha = 0.1$, and an effective charge of $Z_{eff} = 5$, corresponding to an electric field threshold for avalanching of $E_{av}/E_c \approx 2$. As a result, the two sources will be compared for a range of electric fields thresholds up to 250 times the avalanche threshold, which is well above the typical scenarios for tokamak disruptions. The resulting avalanche growth rate is shown in Figs.~\ref{3DPINNgammaAv}(a-b), where the avalanche growth rate using the Rosenbluth-Putvinski source (solid orange curve) and the Chiu-Harvey source (dashed blue curve) are compared to the avalanche growth rates computed with RAMc (black `x' markers), which uses the M\o ller source defined by Eq.~(\ref{Moller}) with a self-consistently computed $f_e \left( p^\prime, \xi^\prime \right)$. It can be seen that the PINN is in good agreement with kinetic simulation for both sources; however, the Chiu-Harvey source is a better predictor of the avalanche growth rate at larger electric fields, where we find that the Rosenbluth-Putvinski source underestimates the avalanche growth rate for scenarios well above avalanche threshold. We note that the present formulation only predicts RE generation, and does not account for the decay of the primary RE distribution, which is due to Eq.~(\ref{dnREdt}) being positive definite. This limitation has recently been overcome in Ref.~\cite{mcdevittpart12025}, where a time dependent solution to the adjoint problem is used to evaluate the decay of the primary distribution of electrons, thus allowing for improved accuracy below threshold. For applications that require RE threshold and decay physics to be accurately described, such as the RE plateau, the primary decay rate inferred in Ref.~\cite{mcdevittpart12025} could be used in conjunction with the present PINN to provide a complete RE avalanche surrogate.

\section{\label{sec:PIREAS}Physical insights from the runaway electron avalanche surrogate}
\subsection{Plasma composition impact on the RE avalanche}
\begin{figure}
\begin{centering}
\subfigure[]{\includegraphics[width=0.33\textwidth]{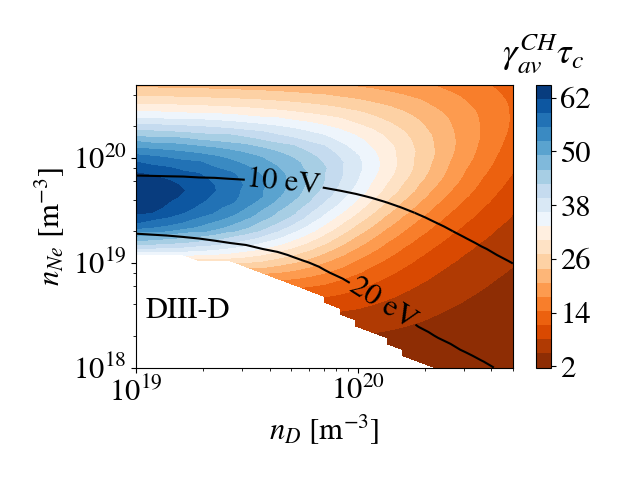}}%
\subfigure[]{\includegraphics[width=0.33\textwidth]{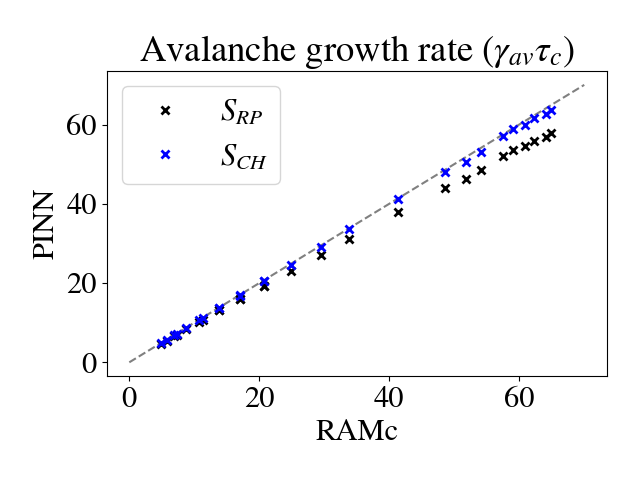}}%
\subfigure[]{\includegraphics[width=0.33\textwidth]{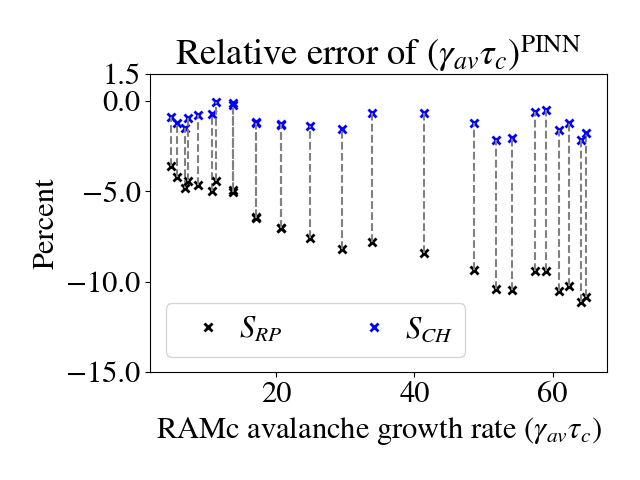}}
\subfigure[]{\includegraphics[width=0.33\textwidth]{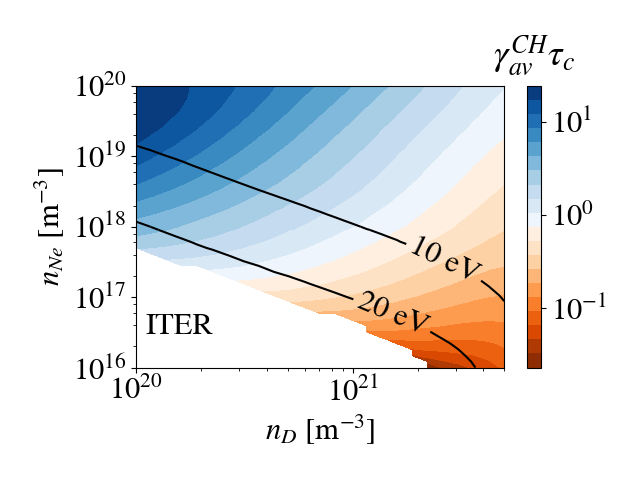}}%
\subfigure[]{\includegraphics[width=0.33\textwidth]{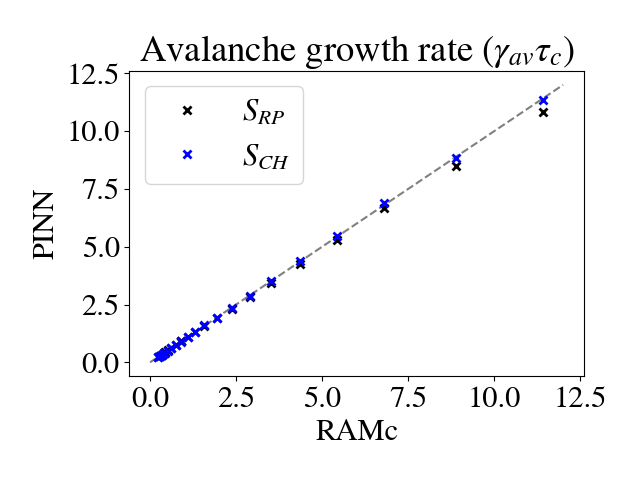}}%
\subfigure[]{\includegraphics[width=0.33\textwidth]{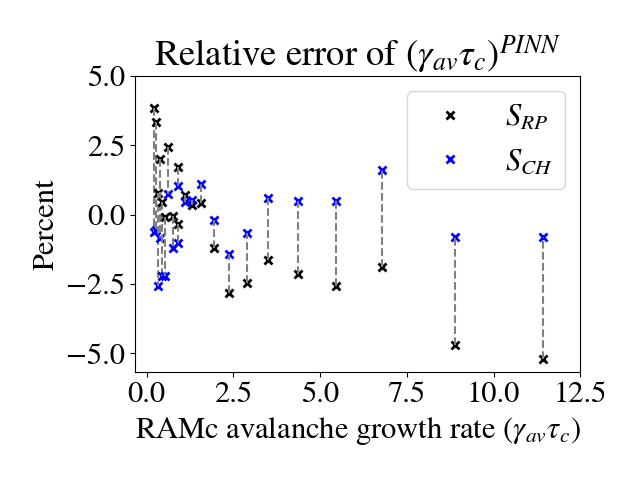}}
\par\end{centering}
\caption{(a,d) The avalanche growth rate using the Chiu-Harvey source, where the black contours correspond to temperatures of 10 and 20 eV. (b,e) The correlation between the PINN predictions of the avalanche growth rate and the true values given by RAMc. (c,f) The relative error between the PINN predictions and the RAMc values of the avalanche growth rate. For panels (b,c,e,f) the avalanche growth rate is computed with the Rosenbluth-Putvinski source (black markers) and the Chiu-Harvey source (blue markers). The top row [panels (a-c)] corresponds to the DIII-D scenario with $j = 2$ MA/m$^2$, and the bottom row [panels (d-f)] corresponds to the ITER scenario (d) with $j = 1$ MA/m$^2$.
}
\label{CRavPINNgammaAv}
\end{figure}

In this section we leverage the PINN described in Sec.~\ref{CRav} to evaluate the avalanche growth rate across a broad range of plasma compositions for the ITER and DIII-D scenarios. To demonstrate the accuracy of the PINN across the regime of avalanching, that is, both in the regime near the avalanche threshold and significantly above the avalanche threshold, we choose $j = 2$ MA/m$^2$ for the DIII-D scenario and $j = 1$ MA/m$^2$ for the ITER scenario. As a result, the electric field will be as large as $E/E_c \approx 1000$ for the DIII-D scenario and as small as $E/E_c \approx 1$ for the ITER scenario [see Figs.~\ref{CRavPINN}(b-c)]. The resulting calculation utilizing the Chiu-Harvey source is shown in Figs.~\ref{CRavPINNgammaAv}(a) and (d). It can be seen that lower deuterium densities [Fig.~\ref{CRavPINNgammaAv}(a)] result in significantly larger $\gamma_{av}^{CH}\tau_c$ compared to the case of larger deuterium densities [Fig.~\ref{CRavPINNgammaAv}(d)]. Specifically, the avalanche growth rate is seen to be peaked in the region of disruption relevant temperatures (black contours, between ten and twenty eV), low deuterium densities, and a plasma composition containing a significant amount of impurities ($n_{Ne} \gtrsim n_D$). At larger deuterium densities [Fig.~\ref{CRavPINNgammaAv}(d)], however, we see that the avalanche growth rate can approach small values when the amount of impurities in the plasma becomes negligibly small ($n_{Ne} \ll n_D$).

Verification of the avalanche growth rate predicted from the PINN is done by selecting cases spanning a broad range of plasma compositions and evaluating the avalanche growth rate with RAMc. Specifically, twenty three plasma compositions along the $T_e = 10$ eV contour are selected for both the ITER and DIII-D scenarios. The resulting comparison between the PINN and RAMc is shown in Figs.~\ref{CRavPINNgammaAv}(b-c) and (e-f), where the predicted avalanche growth rate from the PINN is evaluated for both secondary electron sources $[S_{RP}(p,\xi), S_{CH}(p,\xi)]$. We find excellent agreement between the PINN and RAMc [Figs.~\ref{CRavPINNgammaAv}(b,e)] for the Chiu-Harvey source (blue markers), whereas the Rosenbluth-Putvinski source (black markers) begins to under predict for sufficiently large values of the avalanche growth rate, with deviations particularly evident for the DIII-D scenario. A more quantitative comparison is shown in panels (c) and (f), where the error of the avalanche growth rate predicted by the PINN relative to that computed by RAMc is plotted in percent. For the DIII-D scenario we see that the under predictive behavior of the Rosenbluth-Putvinski source reaches values greater than ten percent at low deuterium densities, whereas the Chiu-Harvey source remains within two and a half percent of the RAMc value of the avalanche growth rate. A similar trend is observed for the ITER scenario. However, at generally larger deuterium densities where the electric field is relatively small, we see that both sources are within five percent accuracy. The Chiu-Harvey source remains accurate within two and a half percent for both the ITER and DIII-D scenarios, demonstrating the high-fidelity predictive capability of the PINN.

\subsection{Dependence of RE avalanche efficiency on plasma composition}
While the RE avalanche surrogate provides a powerful and rapid means of predicting the avalanche growth rate for a broad region of plasma parameters, a convenient quantity of interest is known as the ``avalanche efficiency''. This quantity typically predicts the required drop of poloidal flux for an order of magnitude increase of the RE population. Since the available poloidal flux scales with the plasma current, this provides a convenient means of placing an upper bound on the final amount of RE current that can be generated by the avalanche mechanism during a current quench.  As will be shown in this section, by using the trained PINN described in Sec.~\ref{CRav} to evaluate the rate of RE avalanching $\gamma_{av}$, the efficiency of RE avalanching can be rapidly computed. 

We consider the number of exponentiations of the RE current during a current quench given by $N_{exp} \equiv \int_{t_{i}}^{t_{f}}\mathrm{d}t\gamma_{av}$. If we take the familiar limit where the electric field strength is much larger than the threshold electric field for avalanching, and that the avalanche growth rate scales linearly with the electric field strength~\cite{rosenbluth1997theory}, then we can approximate $\gamma_{av} \approx \gamma_{exp}E/E_c$, where $\gamma_{exp}$ is a constant that characterizes the efficiency of the avalanche growth rate. Noting that the inductive electric field is related to the change of plasma current by $E \approx E_\varphi = L/(2\pi R_0)(d I/d t)$, where $L$ is the plasma self-inductance, and $R_0$ is the major radius of the tokamak, the amount of current for one exponentiation of the RE current can be shown to be
\begin{equation}
N_{exp} \approx \int_{t_{i}}^{t_{f}}\mathrm{d}t \gamma_{exp}\frac{E}{E_c} = \frac{\gamma_{exp} L}{2\pi R_0 E_c} \int_{t_{i}}^{t_{f}}\mathrm{d}t \frac{d I}{d t} = \frac{\gamma_{exp}\mu_0}{2\pi E_c} [I(t_{f}) - I(t_{i})]
, \label{eq:Nexp}
\end{equation}
where we approximated $L\approx \mu_0 R_0$. If we express this result in the more convenient unit of base ten amplifications, this yields $\ln10N_{10} = N_{exp}$. The amount of current for one order of magnitude increase of the RE current $I_{10}$~\cite{boozer2018pivotal} can be evaluated by setting $N_{10} = 1 \Rightarrow N_{exp} = \ln10$, which yields $I_{10} = (\ln10) 2\pi E_c/(\gamma_{exp}\mu_0)$. While we have assumed that several plasma parameters remain constant between $t_i$ and $t_f$, Equation~(\ref{eq:Nexp}) does not assume a constant electric field, where the number of exponentiations of the RE current only depends on the change of plasma current $I(t_{f}) - I(t_{i})$ and $I_{10}$.

While the aforementioned derivation of $I_{10}$ has led to estimations of $\approx$ 0.94 MA for ITER~\cite{boozer2018pivotal}, two key assumptions were made for this estimate. First, the avalanche growth rate was assumed to scale linearly with the electric field. Second, the plasma was assumed to be well above the avalanche threshold. These two assumptions led to the approximation $\gamma_{av} \approx \gamma_{exp}E/E_c$ and allow the integral defining $N_{exp}$ defined in Eq. (\ref{eq:Nexp}) to be performed, thus defining a unique value of $I_{10}$. An additional assumption was that the plasma composition and inductance $L$ were fixed, such that they could be removed from the time integration. Regarding the first assumption, it is well known that the presence of impurities during a tokamak disruption leads to the avalanche exhibiting a \textit{non-linear} dependence on the electric field strength [compare Figs.~\ref{3DPINNgammaAv}(a) and (b)]. As a result, the time integral defining $N_{10}$ no longer can be performed analytically.  To obtain a representative value of the efficiency of the RE avalanche for the general case where the avalanche growth rate does not scale linearly with the electric field, and that the system is not well above threshold, we will need to generalize the conventional derivation of $I_{10}$ outlined above. We proceed to include the non-linear dependence of the electric field on the avalanche growth rate, by directly computing $\gamma_{av}$ from the PINN. We do note, however, that we will maintain the assumption that the plasma parameters including the electric field are constant, implying that $\gamma_{av}$ is constant in time. As a result, the number of base ten amplifications of the RE population is then
\begin{equation}
\ln10N_{10} \equiv N_{exp} = \int_{t_{i}}^{t_{f}} \mathrm{d}t\gamma_{av} =\gamma_{av}\Delta t,
\label{I10a}
\end{equation}
and setting $N_{10} = 1$ yields $\Delta t = \ln10/\gamma_{av}$, where $\Delta t$ corresponds to the time required for the RE population to grow by an order of magnitude. Recalling the previously mentioned induction equation $E = (\mu_0/2\pi)(d I/d t)$, for a constant electric field, we can integrate this equation to yield the current drop during a time $\Delta t$, yielding $\Delta I = 2 \pi E \Delta t/\mu_0$. Substituting Eq.~(\ref{I10a}) with $N_{10}=1$ into this expression then gives the current drop required to increase the RE population by an order of magnitude, i.e.
\begin{equation}
I_{10} = \frac{2\pi\ln10}{\mu_0}\frac{E}{\gamma_{av}} = \frac{I_A\ln10}{2}\frac{E}{E_c}\frac{1}{\gamma_{av}\tau_c},
\label{I10b}
\end{equation}
where we noted the Alfv\'en current is given by $I_A = 4\pi m_ec/(e\mu_0)$. Equation~(\ref{I10b}) reduces to the familiar expression in Ref.~\cite{mcdevitt2019avalanche} if the limit $(E/E_c - 1) \approx E/E_c$ is taken. We re-emphasize that the quantity $I_{10}$ derived here assumes the plasma parameters to be constant during the disruption, but will nevertheless give a representative value of the efficiency of the RE avalanche for a given set of parameters. Here, the PINN described in Sec.~\ref{CRav} can be deployed to evaluate $\gamma_{av}\tau_c$ for a range of post thermal quench plasma scenarios, where we consider the same parameters used in Sec.~\ref{CRav}.
\begin{figure}
\begin{centering}
\subfigure[]{\includegraphics[width=0.5\textwidth]{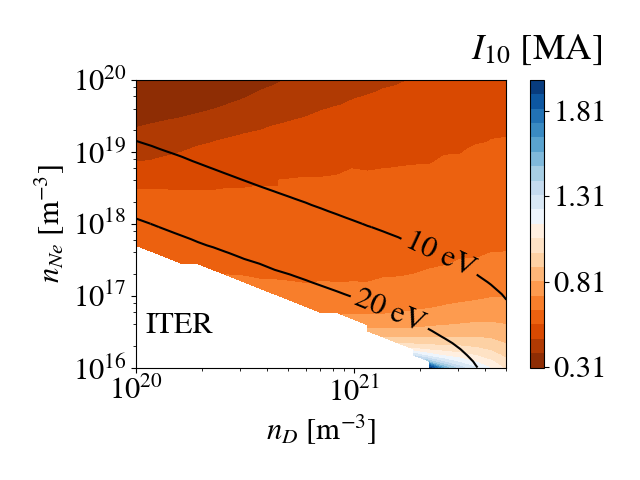}}%
\subfigure[]{\includegraphics[width=0.5\textwidth]{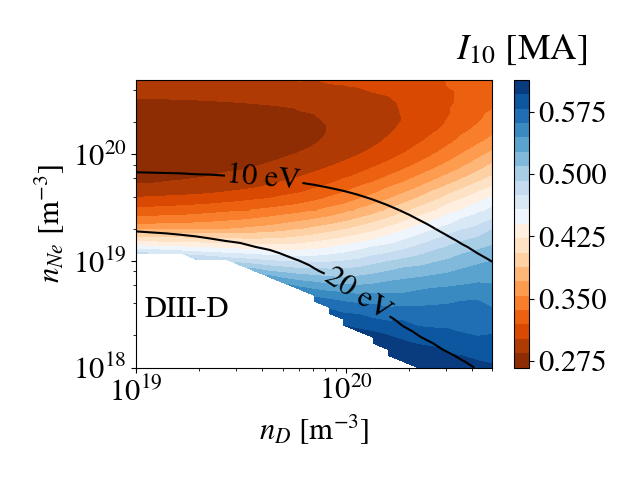}}%
\par\end{centering}
\caption{(a-b) The avalanche efficiency given by Eq.(~\ref{I10b}), where lower values correspond to a larger efficiency, for the ITER (a) and DIII-D (b) scenarios. The PINN utilized is the same as that in Sec.~\ref{CRav}.}
\label{I10scan}
\end{figure}

The resulting calculation of $I_{10}$ for both the ITER and DIII-D scenarios are shown in Fig.~\ref{I10scan}, where the impact of impurities on $I_{10}$ is seen to decrease the avalanche efficiency down to nearly a quarter of a mega ampere, which has been previously reported in Ref.~\cite{arnaud2024impact}. Here, we see that $I_{10}$ is at its minimum, implying a large efficiency, at temperatures smaller than the targeted temperatures for disruptions, unlike the avalanche growth rate; however, in general it can be seen that at low $n_D$ and $n_{Ne} \lesssim n_D$ leads to efficient scenarios for RE avalanche. Another feature shown in Fig.~\ref{I10scan}(a) is the extremely inefficient scenario that is present at large temperatures and large $n_D$, where $I_{10} \gtrsim 1.5$ MA. This region has been seen to result in the plasma approaching the threshold for RE avalanching~\cite{mcdevitt2023constraint}, where here we observe similar trends. We thus anticipate this region of plasma compositions containing a negligible amount of impurities and large deuterium densities to be a possible candidate for the regime that minimizes the amount of RE current generated from RE avalanching.

\section{\label{sec:D}Discussion}
The present work has extended the calculation of the avalanche growth rate using PINNs to the case of partially ionized impurities. As a result, the PINN is able to be utilized as a surrogate for predicting the exponential growth rate of runaway electrons for a broad range of plasma parameters relevant to tokamak disruptions. This paper demonstrates the flexibility of modifying the PINN to account for additional effects that ultimately increase the predictive capability of the surrogate. Moreover, other proof-of-principle surrogates that have been recently developed for describing RE generation mechanisms such as hot-tail generation~\cite{mcdevitt2023physics} and RE decay~\cite{mcdevittpart12025,mcdevittpart22025} can be straightforwardly modified to account for partially ionized impurities, and will be the subject of future work. With regard to the RE avalanche surrogate, a more accurate description of the source of secondary electrons generated by large-angle Coulomb collisions was deployed to increase the fidelity of the surrogate. It was found that the more accurate secondary electron source is required to retain accurate predictions from the surrogate for scenarios where the plasma is significantly above the threshold for RE avalanching. While this secondary electron source requires the computation of a double integral and modestly increases the computational demand for predicting the avalanche growth rate, the general speed of predicting the avalanche growth rate remains in the millisecond level, thus offering a powerful and rapid tool for the larger effort of integrated modeling of a tokamak disruption. 

We also demonstrate an application of integrating an idealized description of plasma power balance into the training of the PINN. While including an additional model typically increases the computational demands for traditional solvers, here it drastically reduces the domain that the PINN needs to train on, thus the performance is significantly improved. We note, however, that both the PINN that integrates a steady-state power balance model and the PINN that learns the PDE over an extremely broad region of plasma parameters are useful, where future work will be to deploy the latter PINN that is parallelized across multiple GPUs and across mini batches of domain space. As a result, this PINN can be utilized in a broader optimization framework. The PINN that embeds the steady-state power balance model demonstrates the path forward in integrated modeling, where future work will be to couple the surrogate presented in this paper with other surrogates describing runaway electrons, thus developing a physics constrained deep learning surrogate for a fluid runaway electron model. One surrogate that was not shown here, but is motivated by the utilization of the RPF against an arbitrary source of electrons, is the generation of REs from nuclear mechanisms such as $\beta$ decay from tritium and Compton scattered electrons from activated material. By utilizing the entire separatrix of the RPF that describes the threshold for electrons running away at a later time, surrogates for nuclear mechanisms will be developed as future work. Finally, the surrogate presented here lacks effects from toroidal geometry, where it is well known that electrons at large minor radii are susceptible to magnetic trapping and will consequently reduce the efficiency of RE avalanching at low collisionalities~\cite{rosenbluth1997theory,nilsson2015kinetic,guo2019toroidal}. While directly accounting for these effects is ongoing work, certain disruption scenarios can lead to negligible effects from toroidal geometry~\cite{mcdevitt2019runaway,arnaud2024impact}, thus leaving the present work applicable in these scenarios. 

\section*{Acknowledgements}
This work was supported by the Department of Energy, Office of Fusion Energy Sciences at the University of Florida under awards DE-SC0024649 and DE-SC0024634, at Los Alamos National Laboratory (LANL) under contract No. 89233218CNA000001, and supported by the U.S. Department of Energy, Office of Science, Office of Workforce Development for Teachers and Scientists, Office of Science Graduate Student Research (SCGSR) program. The SCGSR program is administered by the Oak Ridge Institute for Science and Education for the DOE under contract number DE‐SC0014664. The authors acknowledge the University of Florida Research Computing for providing computational resources that have contributed to the research results reported in this publication. This research used resources of the National Energy Research Scientific Computing Center (NERSC), a Department of Energy Office of Science User Facility using NERSC award FES-ERCAP0032299.

\bibliographystyle{apsrev}
\bibliography{./ref}

\end{document}